\documentclass[10pt,a4paper]{article}
\usepackage[T1]{fontenc}
\usepackage{amsmath,amsfonts, amssymb, amsbsy, mathrsfs, mathtools, bbm, enumitem}
\usepackage[medium]{titlesec}
\usepackage{float}
\usepackage{makeidx}
\usepackage{graphicx}
\usepackage{color}
\usepackage{url}
\usepackage{authblk}
\usepackage{multirow}
\usepackage{lscape}
\usepackage{tensor}
\usepackage{subcaption}
\usepackage[font=small,labelfont=bf]{caption}

\usepackage{mathpazo}
\usepackage[toc]{appendix}
\usepackage[hidelinks,colorlinks=true,linkcolor=red,citecolor=blue,urlcolor=magenta]{hyperref}
\usepackage[left=2.5cm, right=2.5cm, top=3.0cm, bottom=3.0cm]{geometry}

\usepackage{fancyhdr}

\newcommand{\email}[1]{\href{mailto:#1}{#1}}

\newcommand{\df}{\textrm{d}}

\newcommand{\dprime}{{\prime\prime}}

\newcommand{\e}{\textrm{e}}
\newcommand{\hf}{{\frac{1}{2}}}

\renewcommand{\i}{{\rm i}}

\newcommand\scalemath[2]{\scalebox{#1}{\mbox{\ensuremath{\displaystyle #2}}}}

\titleformat{\section}
{\normalfont\large\bfseries}{\thesection}{1em}{}

\titleformat{\subsection}
{\normalfont\normalsize\bfseries}{\thesubsection}{1em}{}

\titleformat{\paragraph}
{\normalfont\normalsize\bfseries}{\theparagraph}{1em}{}

\titleformat{\subparagraph}
{\normalfont\normalsize\bfseries}{\thesubparagraph}{1em}{}

\usepackage[square,numbers,sort&compress]{natbib}

\numberwithin{equation}{section}

\allowdisplaybreaks[1]

\begin{document}
	\setlength{\bibsep}{0pt}
	
	\title{ \textbf{\Large Static Charged Polytropic Spheres with a Cosmological Constant: Physical Acceptability and Trapped Orbits }}
    \author[]{\normalsize Alex Stornelli\thanks{\email{astornelli@smcm.edu}}\ }
    \author[]{\normalsize Anish Agashe\thanks{\email{anagashe@smcm.edu} (corresponding author)}}
	
	\affil[]{{\small \it Department of Physics and Materials Science,
                                
                                St. Mary's College of Maryland,
                                
                                47645 College Dr, St. Mary's City,
                                
                                Maryland, USA 20686}}
	\date{}
	
	\maketitle
	
	\begin{abstract}
            We consider static charged fluid spheres with a cosmological constant. We assume a polytropic equation of state, $p \propto \rho^\Gamma$, and a power law charge distribution, $q\propto r^n$. Using this, we convert the generalised Tolman-Oppenheimer-Volkoff equation into a differential equation for the mass profile. By solving this equation numerically, we analyse both physical and geometric properties of charged polytropic fluid spheres for different values of $n$ and $\Gamma$. By imposing subluminal sound speeds and energy conditions, we restrict ourselves to configurations that are physically acceptable. Then, within these physical models, we study internal trapping of circular geodesics and find the trapping regions in the $n$-$\Gamma$ parameter space. Going beyond the traditionally studied case of null geodesics, we consider orbits of charged and/or massive particles as well. We show that for neutral null particles (and only for them), the possibility of internal trapping is determined purely through geometry. In the other three cases, properties such as the particle's own charge and/or energy also play a role. In general, we find that trapping of all types of particles is allowed for a broad range of $n$ and $\Gamma$.    \\
		
		\noindent {\small \textit{Keywords}: charged polytropic stars, trapped orbits, neutrino trapping, strange stars}
	\end{abstract}

	{
		\hypersetup{linkcolor=black}
		\tableofcontents
	}
	\section{Introduction}
	In the theory of general relativity (GR), static spherically symmetric fluid solutions have been historically significant due to their interesting mathematical properties as well as relevance in modelling compact objects. A common method to build these models is by taking an ansatz on the equation of state (EOS). Within this approach, a frequently used equation of state is the polytropic EOS \cite{tooper1964,pandey1991,nilsson2000,ngub2015}. In this, the pressure ($p$) and density ($\rho$) are related through a simple power law relationship, $p \propto \rho^\Gamma$, where $\Gamma$ is known as the polytropic (or adiabatic) index. Such `polytropic solutions' have also been generalised to include either a charge \cite{ray2003,ray2004,arbanil2013,arbanil2017,noureen2019,acena2024} or a cosmological constant \cite{stuch2016,posada2020,arbanil2020}. However, to the best of our knowledge, there have been no studies that include both. 
    
    Polytropic configurations have been used to describe neutron stars \cite{ferrari2010,alvarez2017,boyle2020,godani2023}, self bound stars (such as compact objects made of quark matter) \cite{lai2008,sunzu2014} and even non-compact objects such as dwarf galaxies and dark matter halos \cite{sanch2021,novo2021}. In analyses of ultra-compact objects using exact solutions like the Tolman VII solution, it is common to study the phenomenon of the internal trapping of circular geodesics  \cite{ishak2001,neary2001,stuch2012,stuch2021,vrba2020,ripple2024}. The possibility of trapping of neutrinos or gravitational waves provides such studies with some astrophysical significance \cite{stuch2021,andrade2001}. Despite polytropes being used to model ultra-compact objects, trapping of geodesics in polytropic configurations has been studied in only a handful of recent papers \cite{novo2017,stuchlik2017,hod2018,hladik2020}.
    
    The goal of this paper is to extend the works mentioned above by providing an exposition to charged polytropic fluid spheres with a cosmological constant and the phenomenon of trapped circular geodesics within such configurations. To solve the polytropic system of equations, it is common to use the ansatz on the density profile due to Tooper \cite{tooper1964} where, $\rho \propto \theta^{\frac{1}{1-\Gamma}}$. This leads to a relativistic analogue of the Lane-Emden equation \cite{tooper1964,stuch2016,novo2017}, which can then be solved. Here, we take a different route and reformulate the generalised Tolman-Oppenheimer-Volkoff (TOV) equation in terms of the mass profile as done previously in \cite{humi}. We extend the treatment in \cite{humi} to include charge and a cosmological constant. This leads to a highly non-linear differential equation with two unknowns -- the mass profile and the charge distribution. We assume a power law charge distribution, $q \propto r^n$, which has been extensively used in the literature \cite{anninos2001,bohm2,arb2015,deb2018,chowdhury2019}. Then, for a range of values of $\Gamma$ and $n$, we solve this differential equation numerically to find the mass profile. Using the mass profile, we calculate the physical properties of the fluid such as density, pressure and sound speed. We also analyse energy conditions and filter out the physically acceptable configurations. After the physical properties, we calculate the geometric properties such as metric functions and spatial curvature for these cases.

    We then turn our attention to the phenomenon of internal trapping of circular geodesics. For a range of values of the central density, $\rho_0$, central pressure, $p_0$, and total charge, $Q$, we explore the $n$-$\Gamma$ parameter space to look for values that allow trapped orbits. We do this for all four possible types of particles: neutral null, neutral timelike, charged null, and charged timelike. Only for the case of neutral null particles (such as photons and gravitons), whether a circular geodesic will be trapped is determined entirely by the gravitational potential. For the other three types, properties such as charge and energy of the particles also play a role. We analyse the effect of these properties on trapping as well. In general, we find that trapping occurs for all the cases for a range of central density and total charge values that correspond to ultra-compact objects as well as a range of $\Gamma$ and $n$. We also find that the bounds on the $n$-$\Gamma$ parameter space for different central densities do not differ in a significant manner but increasing the total charge and the cosmological constant is detrimental to having trapped orbits.    

    The paper is arranged in the following manner: in section \ref{sec-fieldeqs}, we solve the Einstein field equations for charged polytropic fluids and derive the generalised Tolman-Oppenheimer-Volkoff equation. Using this, we present the physical properties of the fluid in section \ref{sec-physprop} as well as geometric properties in section \ref{sec-geoprop}. In section \ref{sec-trapcond}, we derive the generalised effective potential and note down the conditions for internal trapping of various types of particles. We explore different polytropic configurations and check which values of parameters allow for internal trapping in section \ref{sec-paraexp}. Finally, in section \ref{sec-summary}, we summarise our results and provide some concluding remarks.

    We work with geometric units, i.e., $c = 1$, $G = 1$, $\epsilon_0 = 1$, and follow the Landau-Lifshitz space-like convention \cite{mtw}.

	\section{Charged Fluid Spheres with a Cosmological Constant} \label{sec-fieldeqs}
	We use the following line element for a static spherically symmetric space-time,
	\begin{equation}\label{ssmetric}
		\df s^2 = -\e^{2 \Phi(r)}\df t^2 + \e^{2\Psi(r)}\df r^2 + r^2 \left( \df \theta^2 + \sin^2\theta\df\varphi^2 \right)
	\end{equation}
	The time-like unit 4-vector field admitted by such space-time is given by, $u^\alpha = \left[\e^{-\Phi},0,0,0\right] $. The energy-momentum tensor for a perfect fluid is given by,
	\begin{equation}\label{emtensor}
		^{(\rm matter)}{T^\alpha}_\beta  =  (\rho + p) u^\alpha u_\beta + p \delta^\alpha_\beta
	\end{equation}
	where, $\rho(r)$ is the energy density and $p(r)$ is the pressure of the fluid. The fluid is taken to be charged such that it generates a static radial electric field, $E(r)$. The energy-momentum tensor for the electromagnetic field is given by,
	\begin{equation}\label{ememtensor}
		^{(\rm emfield)}{T^\alpha}_\beta = \frac{1}{4\pi} \left(F^{\alpha\sigma}F_{\beta\sigma} - \frac{1}{4}\delta^\alpha_\beta F^{\epsilon\sigma}F_{\epsilon\sigma} \right)
	\end{equation}
	where, the Faraday tensor is given by,
	\begin{equation}\label{faraten}
		F_{\epsilon\sigma} = 2 \delta^1_{[\epsilon}\delta^0_{\sigma]} E(r)
	\end{equation}
	The radial electric field depends on the radial charge distribution in the following manner,
	\begin{equation}\label{chargedist}
		E(r) = \e^{(\Phi + \Psi)} \frac{q(r)}{r^2} 
	\end{equation}

	The Einstein-Maxwell-$\Lambda$ (EM-$\Lambda$) field equations are given by,
	\begin{equation}\label{efelam}
		{R^\alpha}_\beta - \hf R \delta^\alpha_\beta = 8\pi \left(^{(\rm matter)}{T^\alpha}_\beta + ^{(\rm emfield)}{T^\alpha}_\beta\right) - \Lambda \delta^\alpha_\beta
	\end{equation}
	Using equations \eqref{emtensor} - \eqref{chargedist}, the field equations for the metric in \eqref{ssmetric} take the form,
	\begin{subequations}\label{genefe}
		\begin{equation}\label{efe1}
			\frac{\e^{-2\Psi}}{r^2} \left(1 - 2r\Psi^\prime \right) - \frac{1}{r^2} = -8\pi\rho - \frac{q^2}{r^4} -  \Lambda 
		\end{equation}
		
		\begin{equation}\label{efe2}
			\frac{\e^{-2\Psi}}{r^2} \left(1 + 2r \Phi^\prime\right) - \frac{1}{r^2} = 8\pi p - \frac{q^2}{r^4} - \Lambda
		\end{equation}
		\begin{equation}\label{efe3}
			\frac{\e^{-2\Psi}}{r^2}\left[ r\left( \Phi^\prime - \Psi^\prime\right) + r^2\left( \Phi^\dprime + {\Phi^\prime}^2 - \Phi^\prime \Psi^\prime \right)\right] = 8\pi p + \frac{q^2}{r^4} - \Lambda
		\end{equation}
	\end{subequations}
	where, `prime' represent differentiation with respect to the coordinate, $r$. Further, the conservation of energy-momentum leads to only one equation, given by,
	\begin{equation}\label{conseq}
		p^\prime + \Phi^\prime\left(\rho + p\right) = \frac{q q^\prime}{4\pi r^4}
	\end{equation}  
	
	\subsection{Solutions to the Field Equations} \label{sec-efesol}
    Before we present our polytropic models, we will briefly discuss their place in the landscape of static spherically symmetric charged perfect fluid solutions of the EM-$\Lambda$ field equations. Since the system of equations is not closed, additional assumptions are necessary. A classification scheme was presented in \cite{ivanov2002}, based on which variables are specified to close the system. However, we recognise that such assumptions are usually also based on a physical modelling methodology rather than only the purely mathematical issue of closure. For example, we choose to work with a polytropic EOS and a power law charge distribution. Considering this, we identify three broad classes of such modelling methodologies: (I) those that make assumptions on the matter variables (density, pressure, charge, or an EOS); (II) those that make assumptions on the geometric data (one or both metric functions, a relation between metric functions, an embedding or symmetry condition, or a generating function); and (III) those that make assumptions on both the geometry and the matter sector simultaneously. Further, sub-classes can be defined according to the specific assumptions that go in construction of the solutions. 
    
    Here, we will focus on class I since clearly our work falls into this class. Without assuming anything about the metric functions, the first field equation \eqref{efe1} can be simplified as,
	\begin{equation}
		\left[r\left(\e^{-2\Psi} - 1\right)\right]^\prime = -8\pi\rho r^2 - \frac{q^2}{r^2} -  \Lambda r^2 
	\end{equation}
	This can be integrated to give,
	\begin{equation}\label{psieq1}
		\e^{2\Psi(r)} = \left[1 - \frac{2m(r)}{r} - \frac{\epsilon(r)}{r} - \frac{\Lambda r^2}{3}\right]^{-1}
	\end{equation}
	Here, we have defined,
	\begin{subequations}\label{massepdef}
		\begin{equation}\label{massfunc}
			m(r) := 4\pi \int \rho(r) r^2\ \df r
		\end{equation}
		\begin{equation}
			\epsilon(r) := \int \frac{q^2(r)}{r^2}\ \df r
		\end{equation}
	\end{subequations}
	This remains true for any given density profile or charge distribution. Therefore, a general spherically symmetric charged fluid space-time has the following line element,
	\begin{equation}\label{ssle}
		\df s^2 = -\e^{2 \Phi(r)}\df t^2 + \frac{\df r^2}{1 - \frac{2m(r)}{r} - \frac{\epsilon(r)}{r} - \frac{\Lambda r^2}{3}} + r^2 \left( \df \theta^2 + \sin^2\theta\df\varphi^2 \right)
	\end{equation}
	Therefore, $\e^{2\Psi}$ is completely determined through only two matter quantities: $m$ and $q$. Let us see if we can get something similar for $\e^{2\Phi}$. To do this, we use equations \eqref{psieq1} and \eqref{efe2} to write,
	\begin{equation}\label{phieq}
		\Phi^{\prime} = \frac{24\pi p r^3 + 6m + 3\epsilon  - 3r\epsilon^\prime - 2\Lambda r^3}{2r \left( 3r - 6m -3\epsilon - \Lambda r^3\right)}  
	\end{equation}    
    From the existence of pressure on the right hand side above, it is clear that $\e^{2\Phi}$ can only be determined through all three of the matter variables: $m($ (or $\rho$), $p$, and $q$.
    
    To find pressure, we use equation \eqref{phieq} in equation \eqref{conseq} to get,
    \begin{equation}\label{gentov}
        p^\prime + \left\{ \frac{24\pi p r^3 + 6m + 3\epsilon  - 3r\epsilon^\prime - 2\Lambda r^3}{2r \left( 3r - 6m -3\epsilon - \Lambda r^3\right)} \right\}\left( p + \frac{m^\prime}{4\pi r^2} \right) - \frac{q q^\prime}{4\pi r^4} = 0
    \end{equation}
    where, we have also replaced density with mass using, $\rho = \frac{m^\prime}{4\pi r^2}$. This is the generalised Tolman-Oppenheimer-Volkoff (TOV) equation for charged static perfect fluid spheres with a cosmological constant in GR. This is an integro-differential equation in terms of all the three matter quantities\footnote{An alternative would be to use equations \eqref{psieq1} and \eqref{phieq} in equation \eqref{efe3} and obtain a similar integro-differential equation.}. Then, based on the methodology employed to solve this equation, our class I can be divided into three sub-classes where: (Ia) a direct ansatz is taken on two of the three quantities (three further sub-classes: $\left\{\rho, q\right\}$ or $\left\{\rho, p\right\}$ or $\left\{q, p\right\}$); (Ib) a relationship between two of the three quantities is specified and a direct ansatz is taken for the third quantity (three further sub-classes: $\{p(\rho)$, ansatz on $q\}$ or $\{q(\rho)$, ansatz on $p\}$ or $\{p(q)$, ansatz on $\rho\}$); and (Ic) a relationship between all three quantities is specified making them interdependent (three further sub-classes: $\left\{p(\rho), q(\rho)\right\}$ or $\left\{p(\rho), q(p)\right\}$ or $\left\{p(q), q(\rho)\right\}$). 
    
    \subsection{Charged Polytropic Solutions}
    We will proceed to work with class Ib, specifically the subcase of $\{p(\rho)$, ansatz on $q\}$. We begin by choosing a polytropic EOS,
    \begin{equation}\label{polyeos}
        p(\rho) = \kappa\rho^\Gamma
    \end{equation}
    where, $\Gamma$ is a constant known as polytropic index. For a given polytropic index, the constant, $\kappa$, is fixed through initial conditions on pressure and density since we can write,
    \begin{equation}\label{statepara}
        \kappa = \frac{p_0}{\rho_0^\Gamma}
    \end{equation}
    where, $p_0 = p(0)$ and $\rho_0 = \rho(0)$ are the central pressure and central density of the fluid sphere.

    We reformulate the polytropic EOS in terms of mass by writing,    
    \begin{equation}
        p(m) = \kappa \left(\frac{m^\prime}{4\pi r^2} \right)^\Gamma
    \end{equation}
    Using this in the TOV equation, we get,
    \begin{multline}
        \kappa \Gamma \left(\frac{m^\prime}{4\pi r^2} \right)^\Gamma \left(\frac{m^\dprime}{m^\prime} - \frac{2}{r}\right) \\ + \left\{ \frac{24\pi \kappa\rho^\Gamma r^3 + 6m + 3\epsilon  - 3r\epsilon^\prime - 2\Lambda r^3}{2r \left( 3r - 6m -3\epsilon - \Lambda r^3\right)} \right\} \left\{1 + \kappa\left(\frac{m^\prime}{4\pi r^2} \right)^{\Gamma - 1} \right\} \left(\frac{m^\prime}{4\pi r^2} \right) - \frac{q q^\prime}{4\pi r^4} = 0
    \end{multline}
    This is a differential equation with two unknowns: $m$ and $q$. To solve this, one needs an ansatz for one of these quantities such that the other can be solved for. We choose to take the following ansatz on the charge distribution,
    \begin{equation}\label{qans}
        q(r) = Q\frac{r^n}{r_b^n}
    \end{equation}
    where $n$ is real, $r_b$ is the boundary radius of the sphere, and $Q = q(r_b)$ is the total charge of the sphere. Using this, we get a differential equation for the mass profile,
    \begin{multline}\label{masseq}
        r^2 \kappa \Gamma\left[ m^\dprime\left( \frac{m^\prime}{4\pi r^2} \right)^{\Gamma - 1} - 8\pi r \left( \frac{m^\prime}{4\pi r^2} \right)^{\Gamma} \right]\\
        + r^2 m^\prime \left[ 1 + \kappa\left( \frac{m^\prime}{4\pi r^2} \right)^{\Gamma - 1} \right] {\left[ \frac{24\pi r^3 \kappa\left( \frac{m^\prime}{4\pi r^2} \right)^{\Gamma} + 6 m + \frac{6 Q^2(1-n)r^{2 n-1}}{r_b^{2 n}(2 n - 1)} - 2\Lambda r^3}{6 r^2 - 12 m - \frac{6 Q^2 r^{2 n-1}}{r_b^{2 n}(2 n - 1)} - 2\Lambda r^4}  \right]}\\ - \frac{Q^2 n r^{2 n - 1}}{r_b^{2 n}} = 0
    \end{multline}
    This equation will serve as a `master equation' for the analysis in this paper\footnote{An alternative would have been to impose an ansatz on the mass (or density) profile and derive a similar master equation in terms of charge distribution. }. If we can solve this equation to find $m(r)$, then all the other physical and geometric properties of the corresponding polytropic configuration will follow from there. However, we can only do this numerically owing to the highly non-linear nature of equation \eqref{masseq}. We use numerical integrators and differential equation solvers in Python \cite{scipy2020} to obtain the mass profiles. The initial conditions used are, $m(r_i) = 1\times 10^{-12}$ and $m^\prime(r_i) = \frac{\rho_0}{4\pi r_i^2}$, where, $r_i = 0.001$. The initial values, $m(r_i)$ and $r_i$, are taken to be not exactly zero to avoid numerical errors that may arise due to division by zero.
    
    We consider the range, $1\le \Gamma \le 5$, since the polytropic index for compact objects falls in this range \cite{douchin2001,haensel2002}. We note that the above equation becomes singular at centre for $n < \hf$ and everywhere for $n = \hf$. To avoid this, we choose the range, $1\le n \le 5$. We also vary the values of $\rho_0$ and $Q$ and present several cases. The values of these two parameters are taken to be their typical values for ultra-compact objects, which in geometric units become, $\rho_0 \sim 10^{-9}$ m$^{-2}$ and $Q \sim 10^3$ m \cite{ray2003,ray2004} while $\kappa$ is calculated by fixing the ratio, $\frac{p_0}{\rho_0} = 0.2$. We use this value because it satisfies the constraint, $\frac{p_0}{\rho_0} < \frac{1}{\Gamma}$ that comes from causality \cite{stuch2016,stuchlik2017}. Further, the cosmological constant is fixed by cosmology \cite{planck2018} while we take $r_b = 10^4$ m which is a typical size of a neutron star.
    
    We plot the mass profile for different values of these parameter in figure \ref{fig-massprofile} below. We find that for these ranges of different parameters, the mass distribution is relatively well-behaved except for $\rho_0 \gtrsim 10^{-8} $ and $Q \gtrsim 10^4$ as can be seen in the bottom row of the plot. Corresponding to different values of $Q$, the generalised Buchdahl limit \cite{andrea2012} on the total inertial mass can be calculated to be between $4\times 10^3$ m to $6\times 10^3$ m. The total mass, $m(r_b)$, falls in this range for all the well-behaved cases. Further, for this range of the total mass and the total charge, the static radius ($r_s$) (also called the turnaround radius) \cite{german2024} can be calculated to be in the range, $4.2\times 10^{18}\ {\rm m} < r_s < 5.6 \times 10^{18}$ m. As expected, the size of the polytropic fluid spheres ($r_b$) that we have used here is much smaller than the static radius. This is not surprising and has been already noted for uncharged polytropic configurations, for example, in \cite{stuch2016}. We will show in the proceeding sections that these cases also follow the standard physical acceptability criteria.

    \begin{figure}[H]
        \centering
        \includegraphics[width=\linewidth]{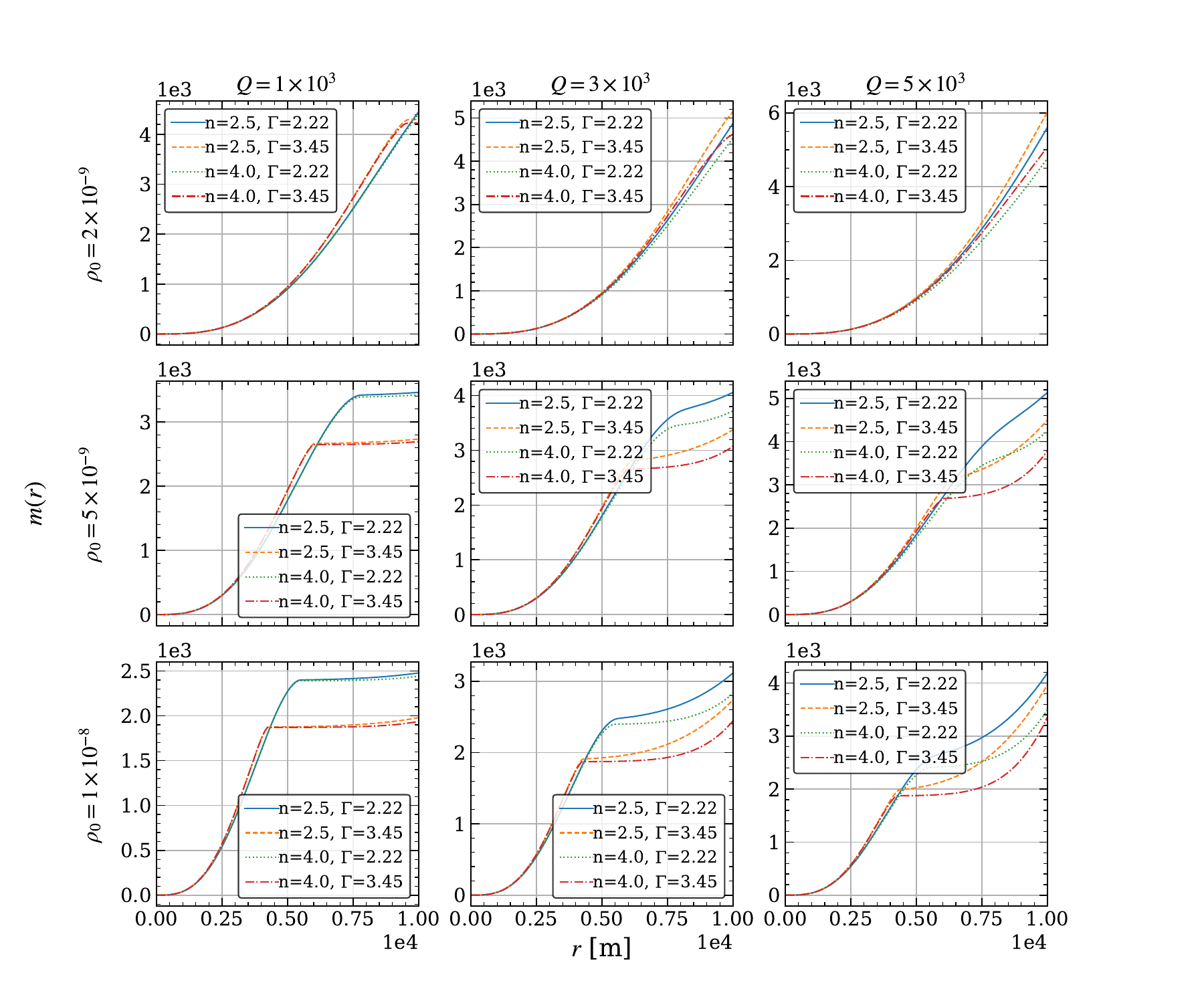}
        \caption{The mass profile, $m(r)$, of various charged polytropic configurations (different values of $n$ and $\Gamma$) obtained by solving equation \eqref{masseq}. The central density, $\rho_0$, and total charge, $Q$, correspond to their typical values for ultra-compact objects \cite{mathews1997,ray2003,ray2004}. The boundary radius is, $r_b = 10^4$ m and $\Lambda = 10^{-52}$ m$^{-2}$. }
        \label{fig-massprofile}
    \end{figure}

    \section{Physical Properties of the Fluid} \label{sec-physprop}

    \subsection{Density and Pressure}
    Once we solve for the mass profile, it is straightforward to calculate the physical properties of the fluid for a given polytropic configuration. For example, the density profile can be found by inverting equation \eqref{massfunc}. This gives,
    \begin{equation}
        \rho(r) = \frac{m^\prime(r)}{4\pi r^2}
    \end{equation}
    The pressure can be found by using the density profile above in equation \eqref{polyeos}. We plot the radial profile of the energy density and pressure for different values of $\Gamma$ and $n$ in figures \ref{fig-denprofile} and \ref{fig-pressprofile} below. For most of the values where $1\le n,\Gamma \le 5$, both the density and pressure are monotonically decreasing and the pressure tends to vanish at the boundary. However, not all configurations exhibit these features. We filter such cases out, along with others that do not follow additional physical acceptability criteria in the following sections.

    \begin{figure}[H]
        \centering
        \includegraphics[width=0.95\linewidth]{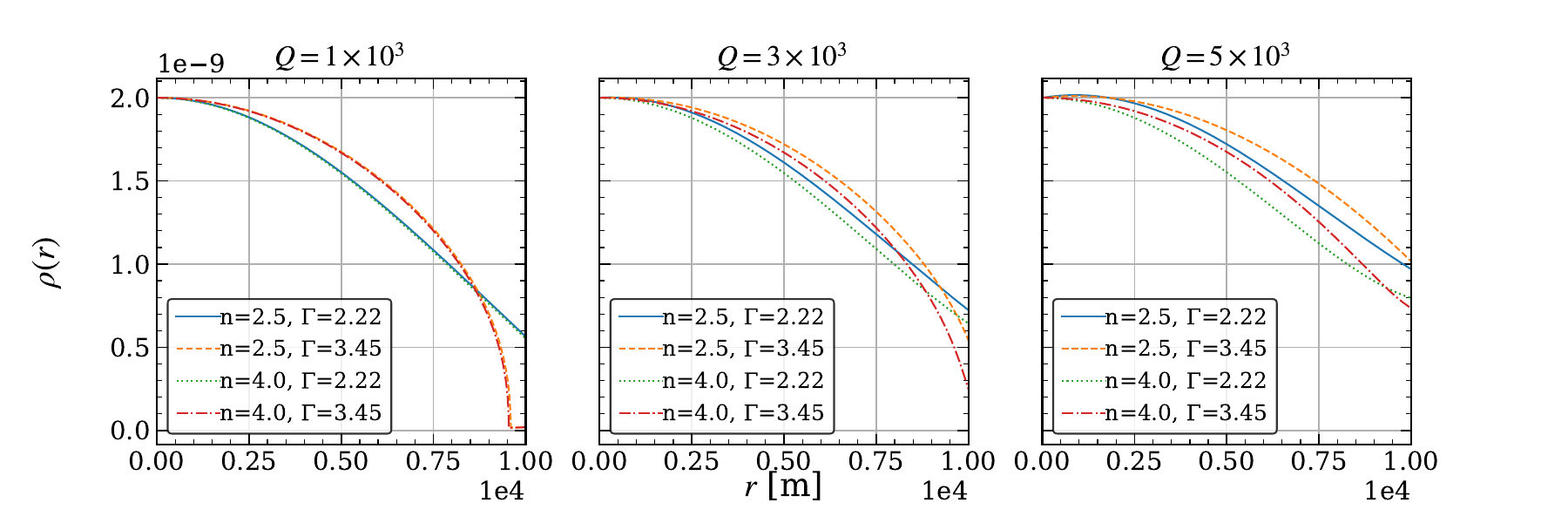}
        \caption{The density profile, $\rho(r)$, for various polytropic configurations. We present the cases where the energy density decreases monotonically throughout the extent of the fluid sphere. }
        \label{fig-denprofile}
    \end{figure}

    \begin{figure}[H]
        \centering
        \includegraphics[width=\linewidth]{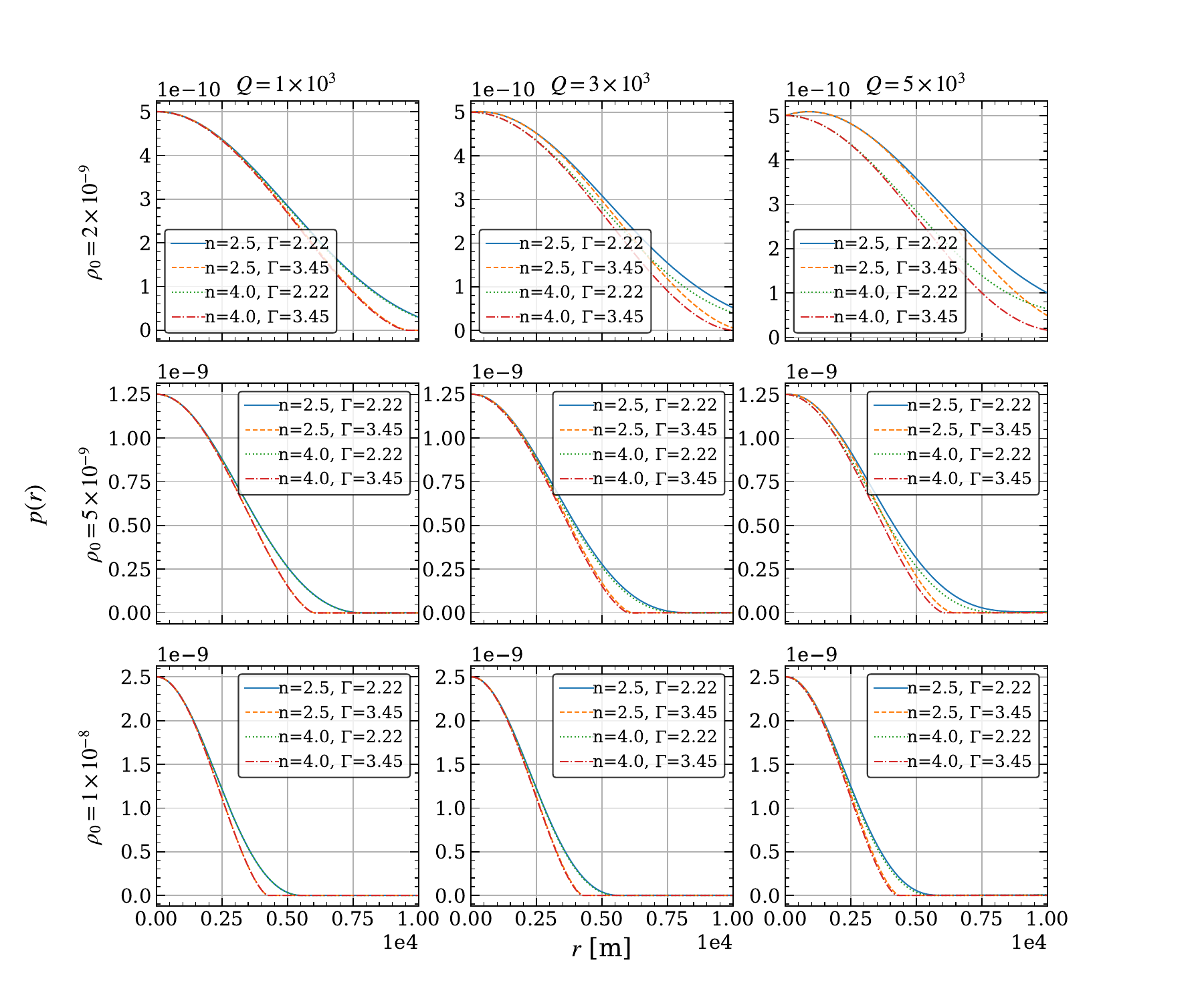}
        \caption{The radial profile for pressure, $p(r)$, for various polytropic configurations. We present the cases where the pressure decreases monotonically throughout the extent of the fluid sphere.  }
        \label{fig-pressprofile}
    \end{figure}

    \subsection{Sound Speed and Bulk Modulus}
    Using equations \eqref{polyeos} and \eqref{statepara}, we find an expression for the speed of sound in the polytropic fluid to be,
    \begin{equation}
        c_s^2(r) = \frac{\partial p}{\partial \rho} = \Gamma\frac{p}{\rho} = \Gamma\frac{p_0}{\rho_0^\Gamma}\left(\frac{m^\prime}{4\pi r^2}\right)^{\Gamma - 1}
    \end{equation}
    The second equality above ensures that if causality is respected (i.e., $c_s^2 < 1$), then the fluid is also a Newtonian fluid since for $\Gamma > 1$. Moreover, at $r = 0$, we must have, $c_s^2(0) < 1$, which fixes the ratio of the central pressure and density as, $\frac{p_0}{\rho_0} < \frac{1}{\Gamma}$. This has already been noted in \cite{stuch2016,stuchlik2017}.
    
    Further, the bulk modulus, $B$, of a polytropic fluid would be given by,
    \begin{equation}
        B(r) = \rho c_s^2 = \Gamma p = \Gamma\frac{ p_0}{\rho_0^\Gamma} \left(\frac{m^\prime}{4\pi r^2}\right)^\Gamma
    \end{equation}
    We plot the sound speed and the bulk modulus in figures \ref{fig-sspeed} and \ref{fig-bulkmod} below. We find that causality is respected in most of the cases as well as the sound speed is monotonically decreasing.

    \begin{figure}[H]
        \centering
        \includegraphics[width=\linewidth]{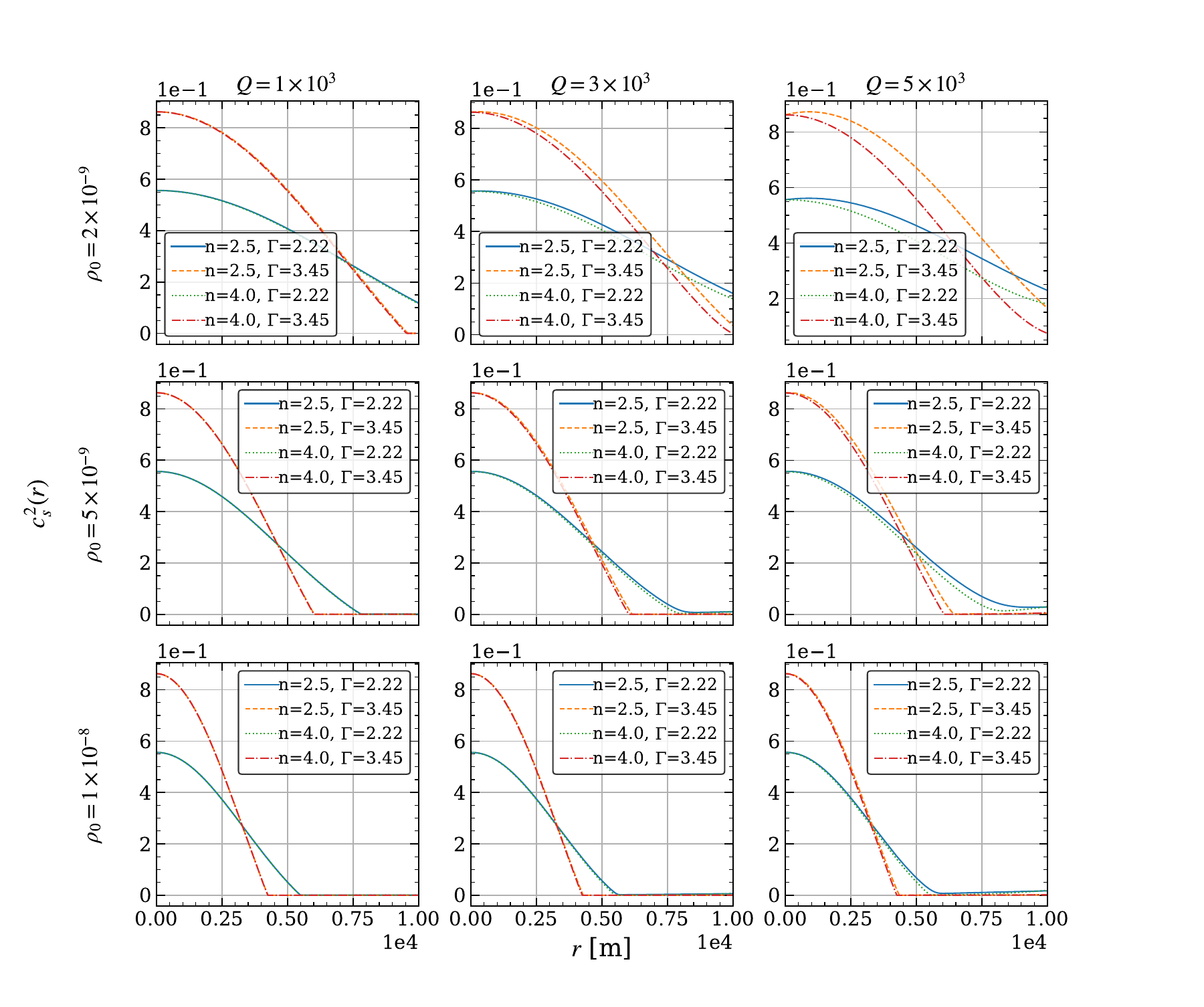}
        \caption{The speed of sound squared, $c_s^2(r)$, as a function of the radial coordinate, $r$. For all the presented cases, we find subluminal sound speeds ($c_s^2 < 1$), i.e., causality is not violated.}
        \label{fig-sspeed}
    \end{figure}

    \begin{figure}[H]
        \centering
        \includegraphics[width=\linewidth]{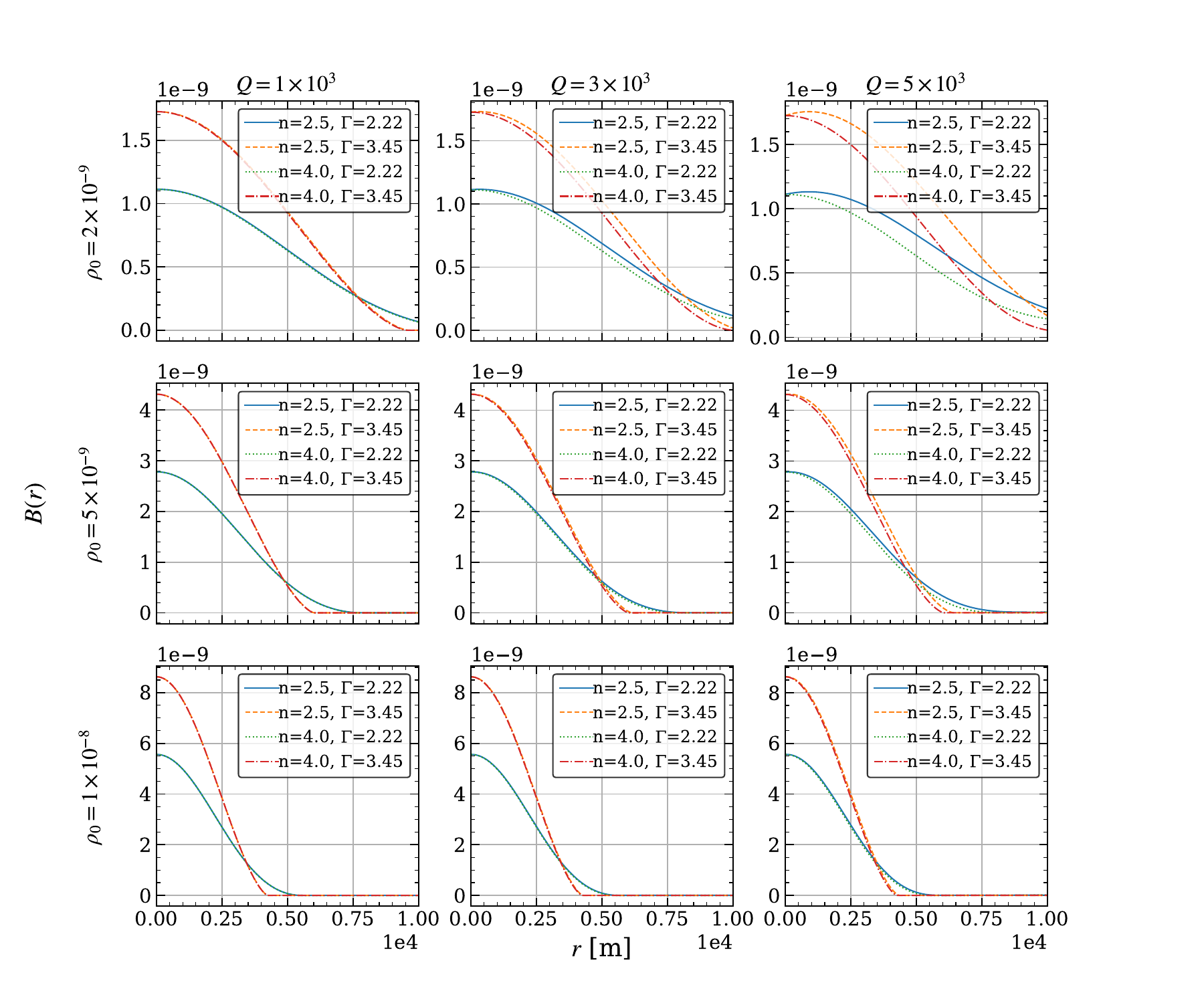}
        \caption{The bulk modulus, $B(r)$, as a function of $r$ for various charged polytropic configurations.}
        \label{fig-bulkmod}
    \end{figure}

    \subsection{Energy Conditions}
\begin{figure}[H]
        \centering
        \includegraphics[width=\linewidth]{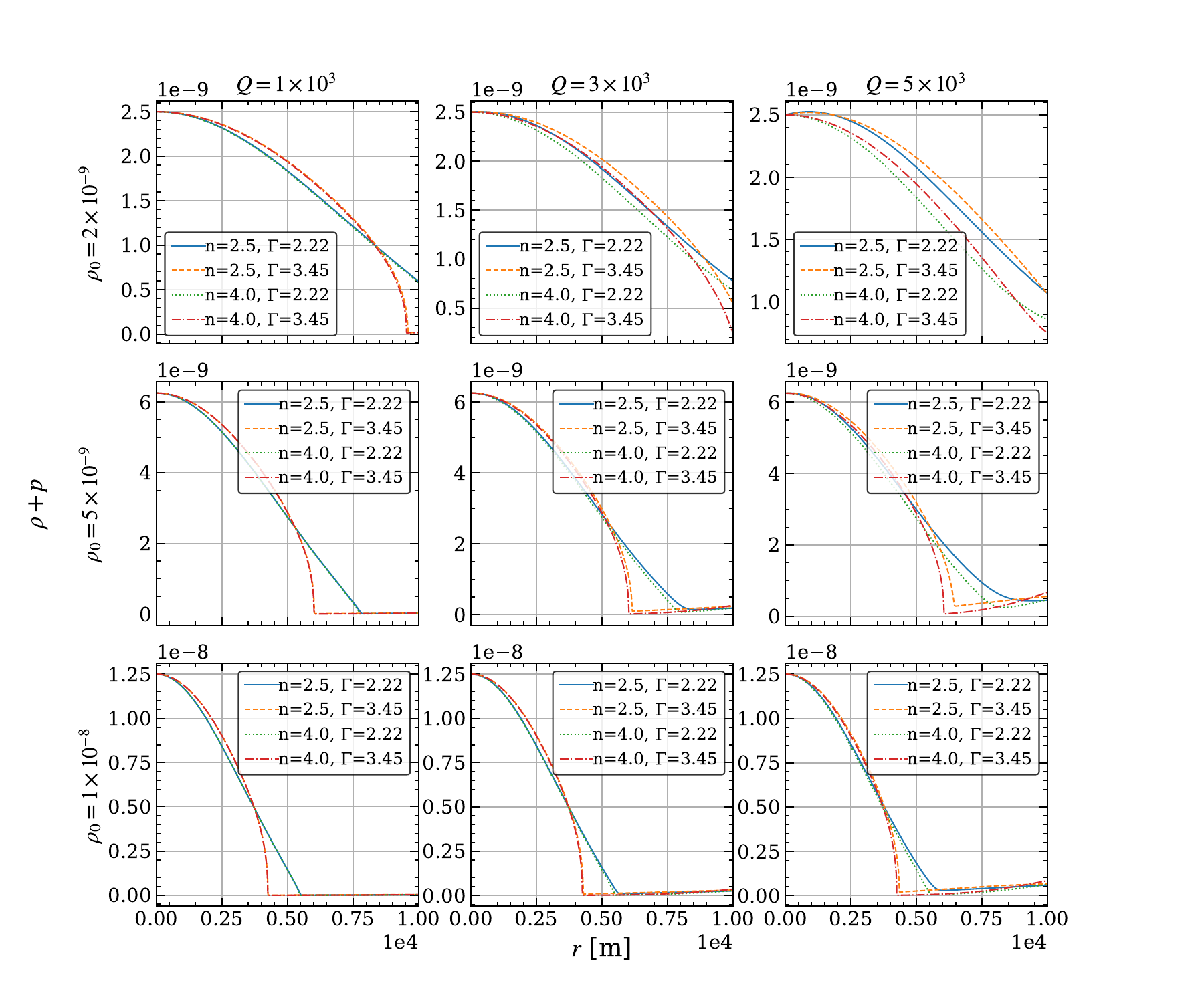}
        \caption{The left hand side of the null energy condition for various charged polytropic configurations. The null energy condition is satisfied for most of the configurations.}
        \label{fig-nec}
    \end{figure}

\begin{figure}[H]
        \centering
        \includegraphics[width=\linewidth]{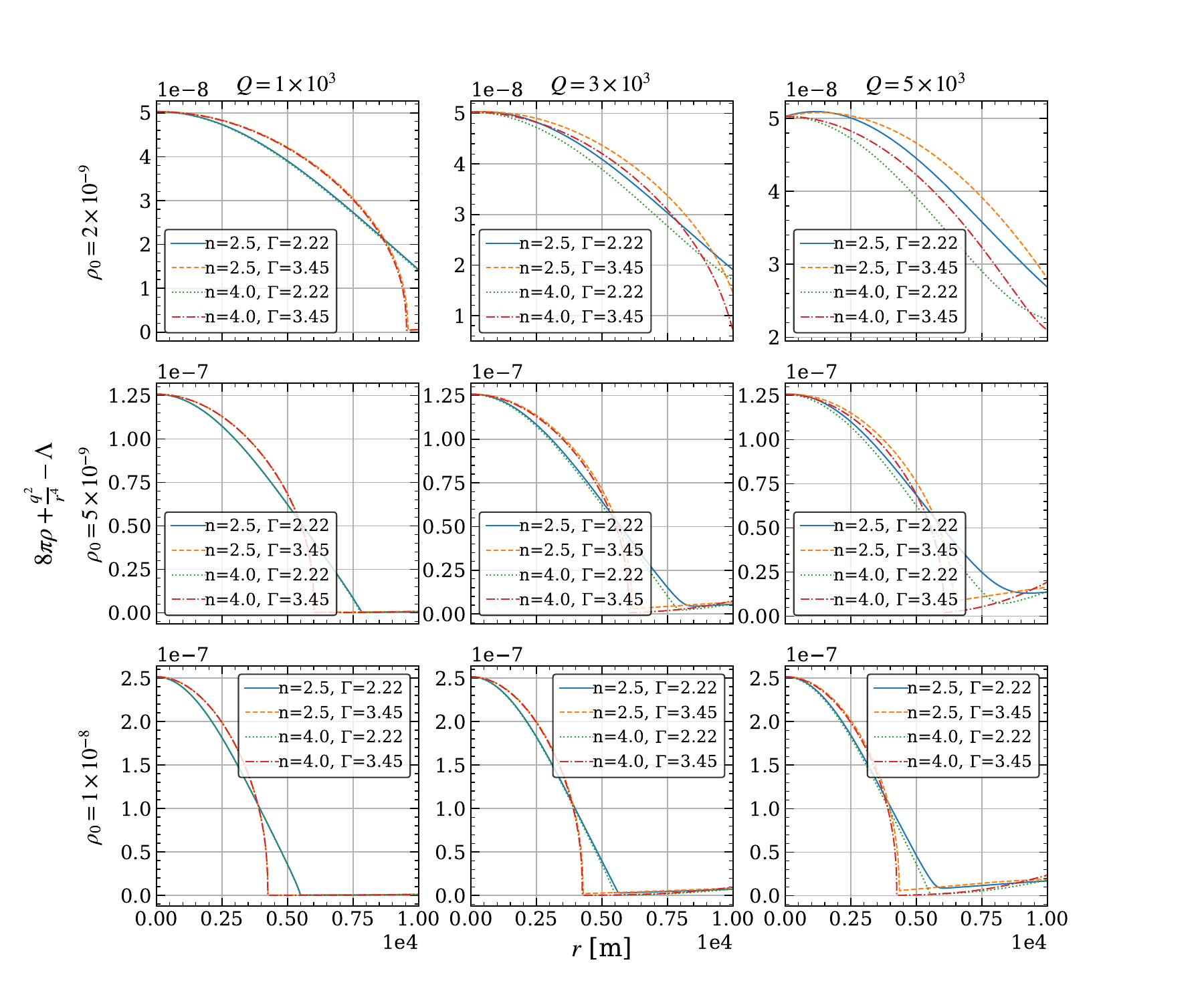}
        \caption{The left hand side of the weak energy condition for various charged polytropic configurations. The weak energy condition is satisfied for most of the configurations. }
        \label{fig-wec}
    \end{figure}

\begin{figure}[H]
        \centering
        \includegraphics[width=\linewidth]{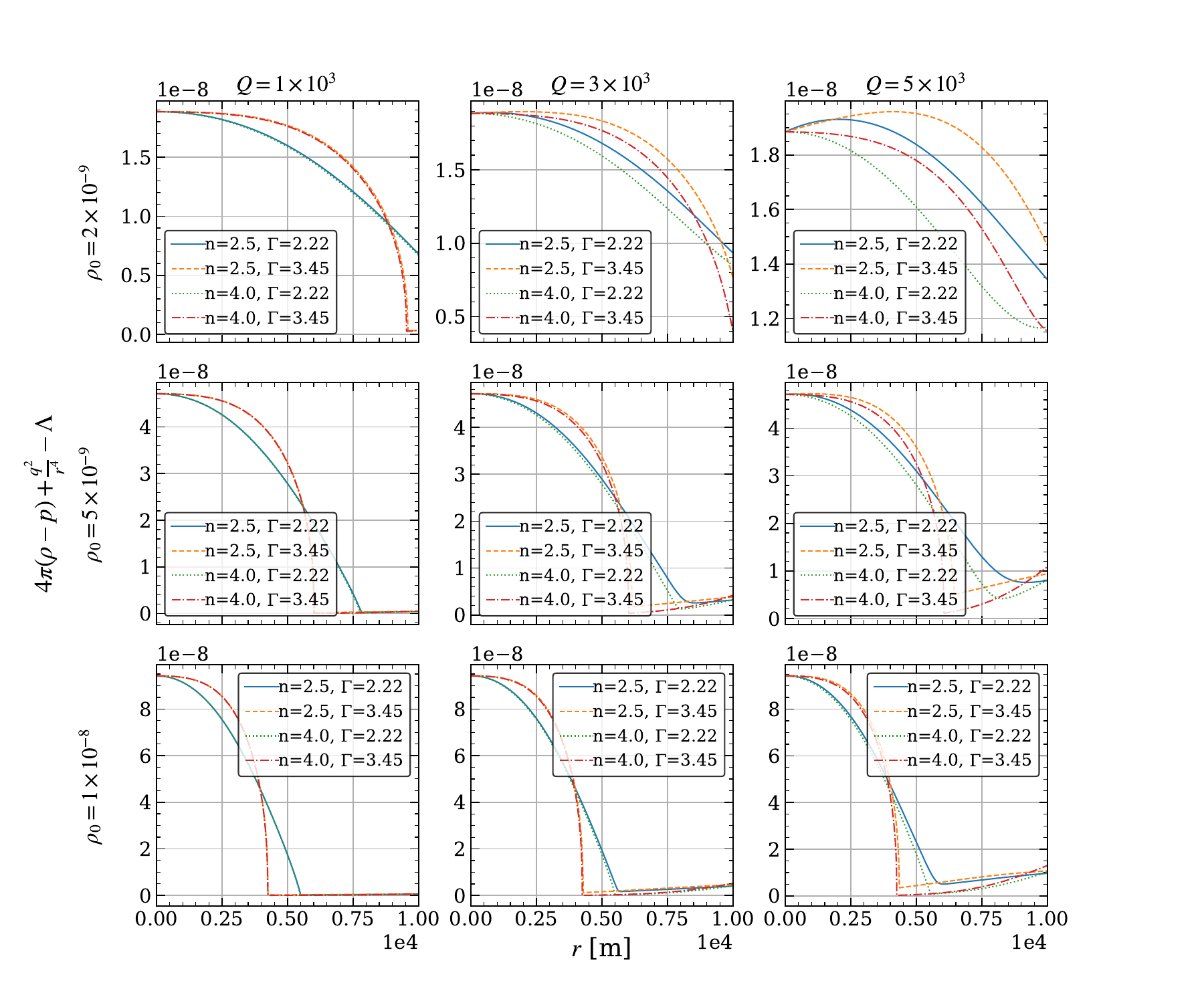}
        \caption{The left hand side of the dominant energy condition for various charged polytropic configurations. The dominant energy condition is satisfied for most of the configurations.}
        \label{fig-dec}
    \end{figure}

\begin{figure}[H]
        \centering
        \includegraphics[width=\linewidth]{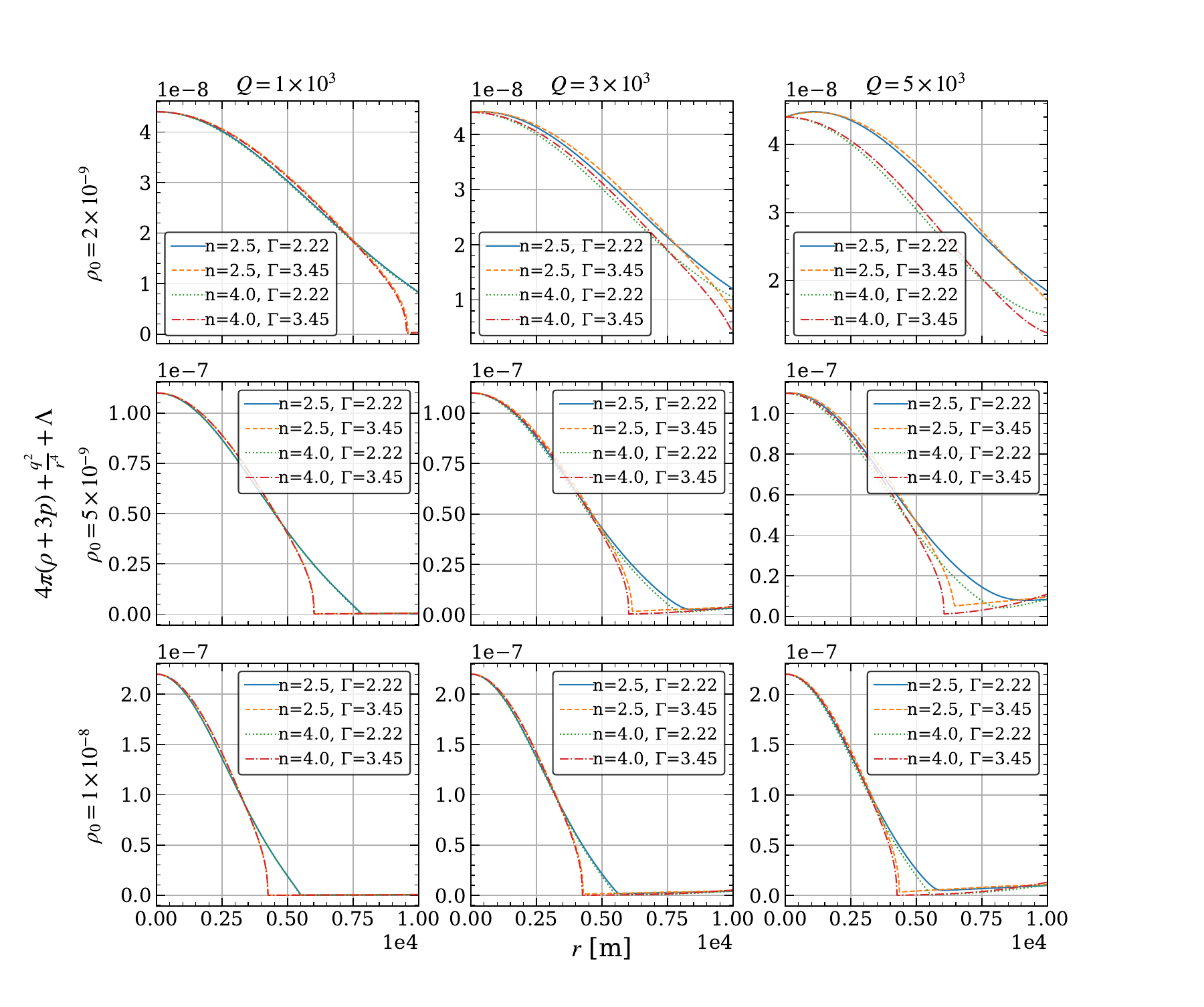}
        \caption{The left hand side of the strong energy condition for various charged polytropic configurations. The strong energy condition is satisfied for most of the configurations. }
        \label{fig-sec}
    \end{figure}

    Using the sound speed, we filter out the configurations (i.e., values of $\Gamma$ and $n$) for which causality is respected. To investigate the physicality of the remaining cases even further, we plot the so-called energy conditions. These are restrictions on the matter content and constitute four individual inequalities as listed below \cite{curiel2014, ripple2024}.
    \begin{align}
        \text{Null Energy Condition}: \quad & \rho + p \ge 0 \\
        \text{Weak Energy Condition}: \quad & 8\pi \rho + \frac{q^2}{r^4} + \Lambda \ge 0\\
        \text{Dominant Energy Condition}: \quad & 4\pi (\rho - p) + \frac{q^2}{r^4} + \Lambda \ge 0 \\
        \text{Strong Energy Condition}: \quad & 4\pi (\rho + 3p) + \frac{q^2}{r^4} - \Lambda \ge 0
    \end{align}
    We plot these energy conditions in figures \ref{fig-nec}, \ref{fig-wec}, \ref{fig-dec}, and \ref{fig-sec}, respectively.

    Using all the figures in this section, we find that most of our models satisfy the basic physical acceptability criteria. We filter out the regions in the $n$-$\Gamma$ parameter space that allow for such configurations and plot the regions of physical acceptability in figure \ref{fig-physaccept} below. We find that there are only very few physically acceptable cases for $\Gamma <2$ for all values of $n$. Moreover, increasing the amount of total charge is detrimental to physical acceptability and as $n$ and $\Gamma$ increase with $Q$, the acceptable region almost vanishes (see the top right plot in figure \ref{fig-physaccept}).  
    \begin{figure}[H]
        \centering
        \includegraphics[width=\linewidth]{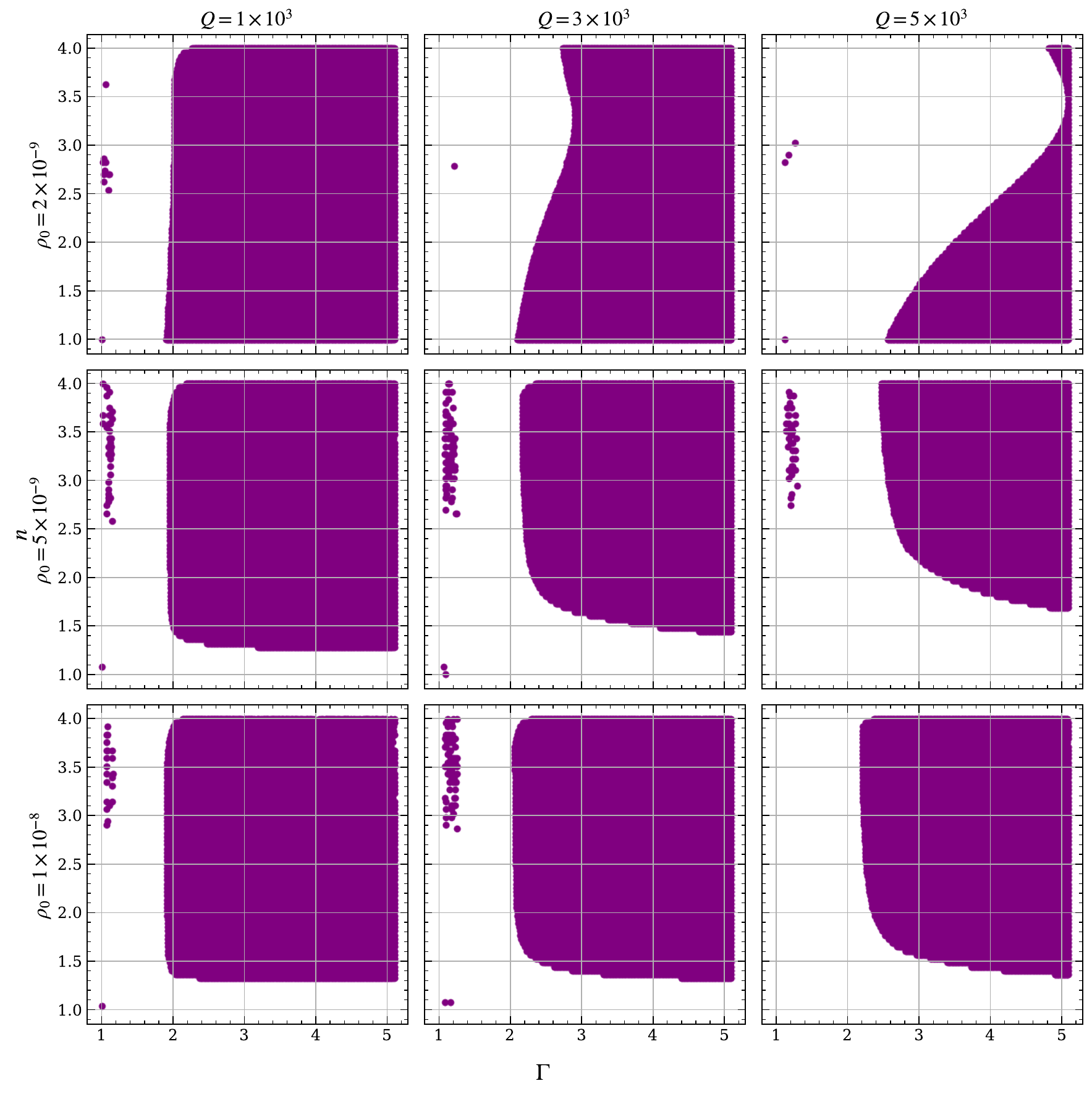}
        \caption{The physically acceptable regions in the $n$-$\Gamma$ parameter space for a given value of central density, $\rho_0$ and total charge, $Q$.}
        \label{fig-physaccept}
    \end{figure}

    \section{Geometric Properties of the Space-time} \label{sec-geoprop}

    \subsection{Metric Functions}
    Using the mass profile and the charge distribution, the metric functions can be found using equations \eqref{psieq1} and \eqref{phieq}. We plot them below in figures \ref{fig-ephiprofile} and \ref{fig-epsiprofile}. We find that the metric coefficients remain regular and finite in most of the cases. However, as the total charge and the central density increases, the metric functions for some configurations do not remain smooth.  

    \begin{figure}[H]
        \centering
        \includegraphics[width=0.92\linewidth]{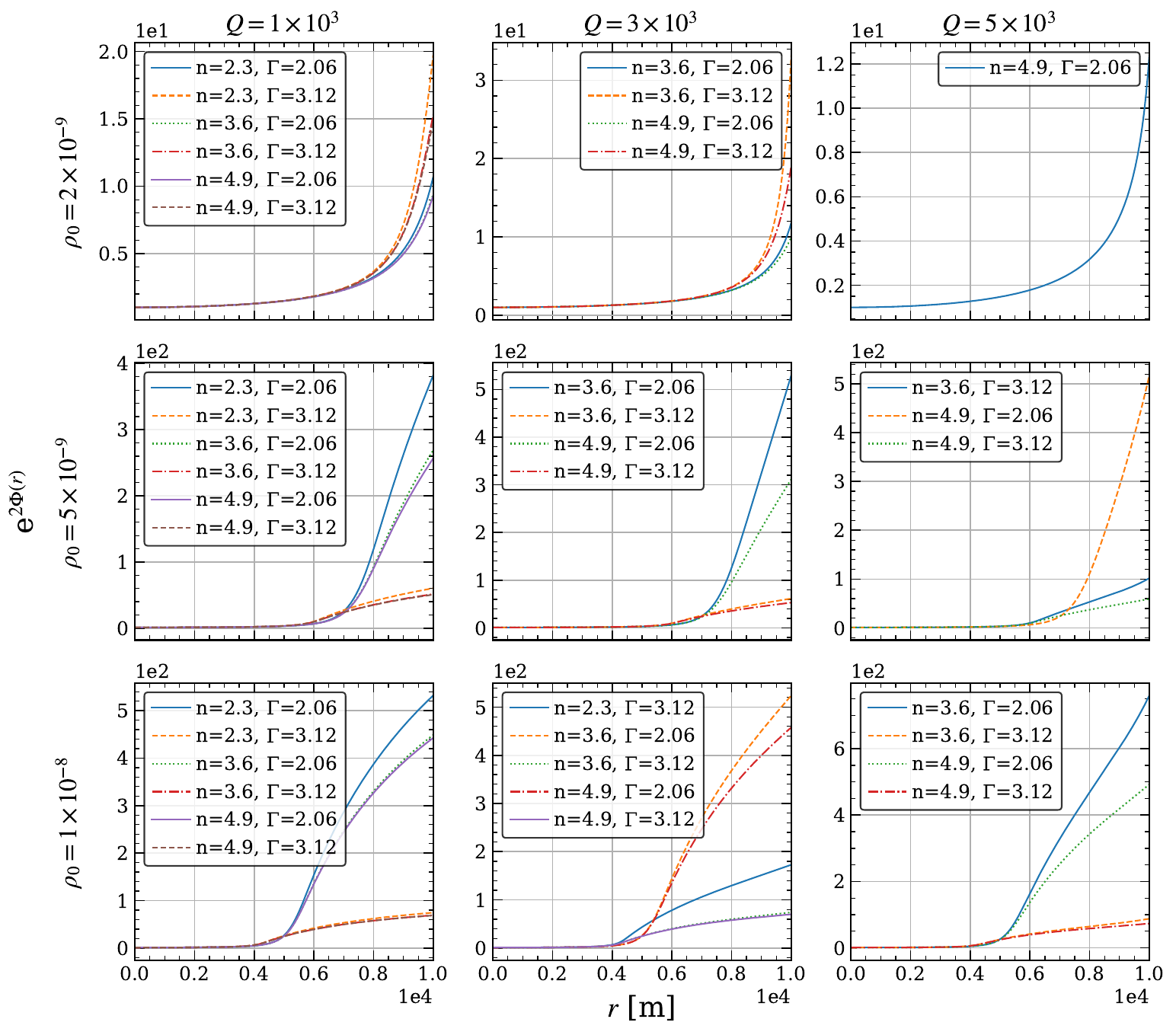}
        \caption{The radial profile of the metric function, $\e^{2\Phi(r)}$. The function remains smooth for most typical values of $\rho_0$ and $Q$. From the bottom row, one can see that for large values ($\rho_0 > 10^{-8}$ and $Q>10^4$), discontinuities may be present in the metric function.}
        \label{fig-ephiprofile}
    \end{figure}

    \begin{figure}[H]
        \centering
        \includegraphics[width=0.92\linewidth]{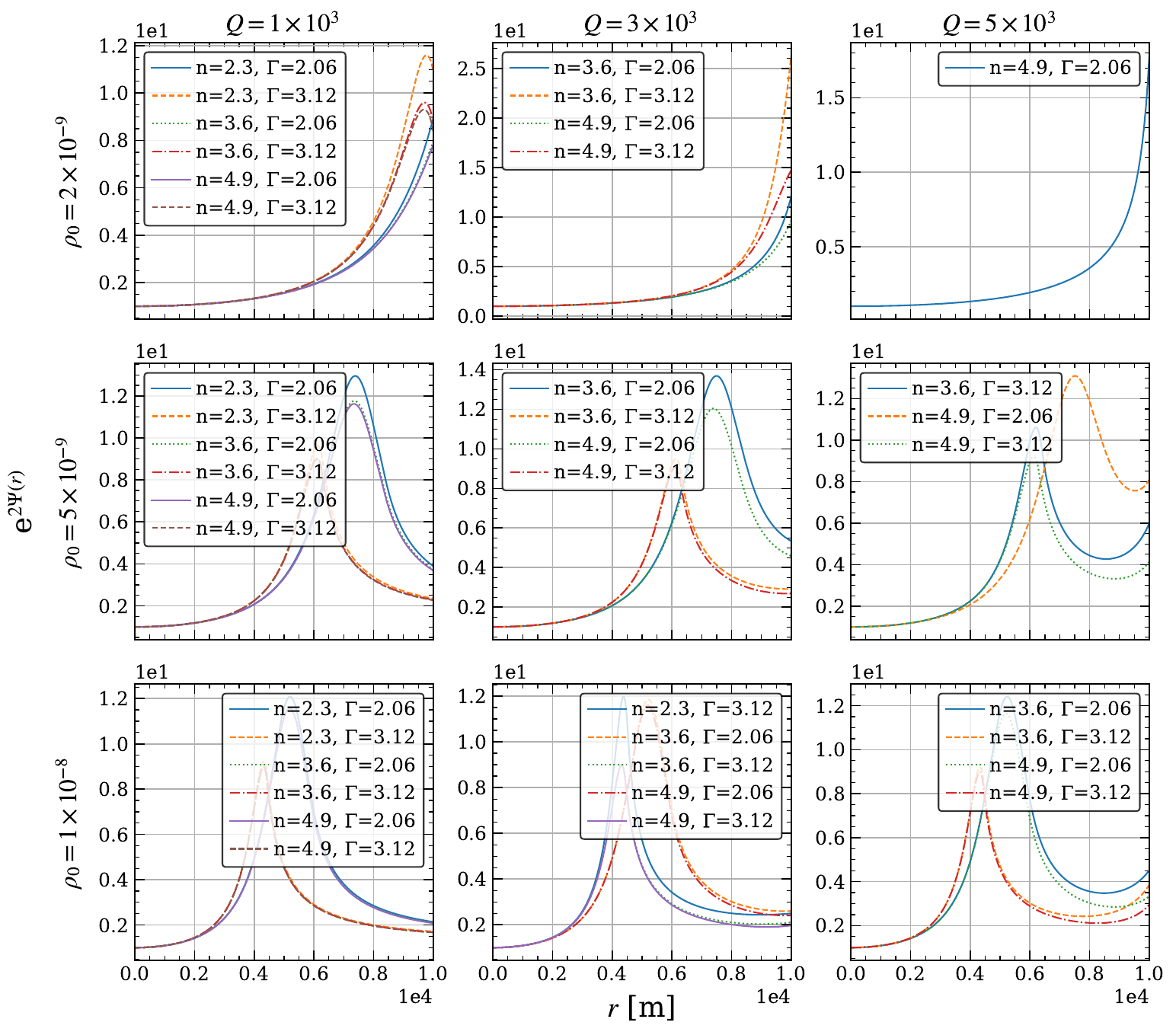}
        \caption{The radial profile of the metric function, $\e^{2\Psi(r)}$. The function remains smooth for most typical values of $\rho_0$ and $Q$.}
        \label{fig-epsiprofile}
    \end{figure}
    
    \subsection{Spatial Curvature}
\begin{figure}[H]
        \centering
        \includegraphics[width=0.92\linewidth]{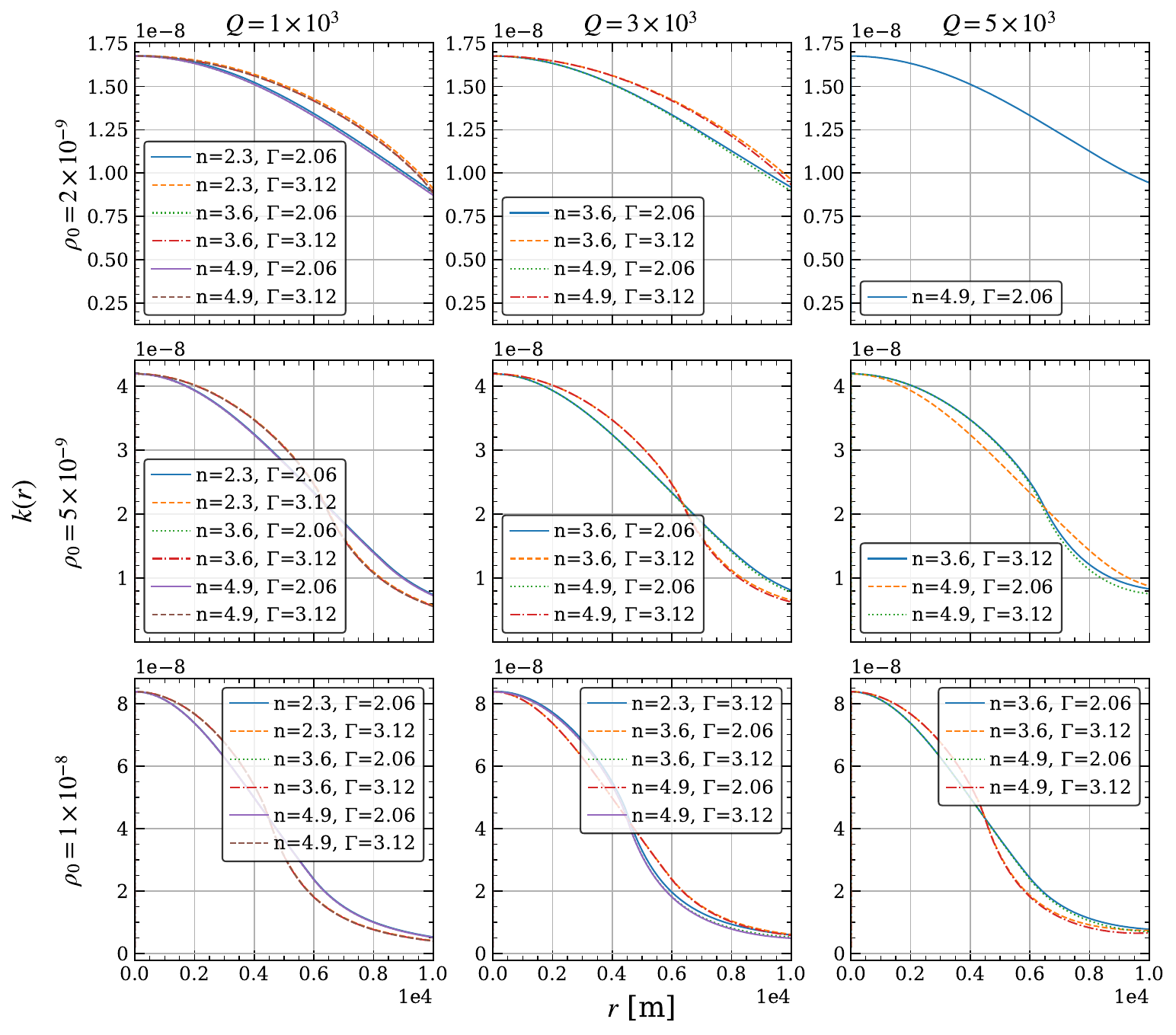}
        \caption{The radial profile of the spatial curvature, $k(r)$, for various polytropic configurations. The spatial curvature remains positive, i.e., the spatial geometry remains spherical for all of our models ($1\le n,\Gamma\le 5$) constructed with realistic values of $\rho_0$ and $Q$.}
        \label{fig-spatialprofile}
    \end{figure}

    The line element for a spherically symmetric charged perfect fluid space-time (equation \eqref{ssle}) can be written in the form,
    \begin{equation}
        \df s^2 = -\e^{2\Phi}\df t^2 + \frac{\df r^2}{1 - k(r) r^2} + r^2\left( \df \theta^2 + \sin^2\theta \df \varphi^2 \right)   
    \end{equation}
    where,
    \begin{equation}
        k(r) := \frac{1 - \e^{-2\Psi(r)}}{r^2} = \frac{2m(r) + \epsilon(r)}{r^3} + \frac{\Lambda}{3} 
    \end{equation}
    The function, $k(r)$, above is directly proportional to the scalar curvature of the spatial ($t = {\rm const.}$) hypersurfaces. In figure \ref{fig-spatialprofile}, we plot the spatial curvature as a function of the radial coordinate, $r$. We find that for our chosen values of $\rho_0$ and $Q$, the spatial geometry remains spherical (i.e., $k(r) > 0$) throughout the extent of the fluid sphere for $1 \le n,\Gamma \le 5$. However, in the case of a negative cosmological constant, the spatial curvature may take negative values but only when its value is comparable to the central density ($|\Lambda| \sim \rho_0$).

    \section{Effective Potential and Trapped Circular Geodesics} \label{sec-trapcond}
    The equations of motion for a test particle in a space-time containing charge are given by,
    \begin{equation}
        \ddot{x}^\alpha + {\Gamma^\alpha}_{\beta \sigma}\dot{x}^\beta\dot{x}^\sigma - e {F^\alpha}_\beta \dot{x}^\beta = 0
    \end{equation}
    where, `dot' represents a differentiation with respect to some affine parameter, $\boldsymbol{\Gamma}$ is the Levi-Civita connection, $e$ is the specific charge (charge per unit mass) of the test particle, and $\boldsymbol{F}$ is the electromagnetic field tensor (also known as Faraday tensor). For electrically neutral test particles, $e = 0$, and the equation of motion is reduces to the usual geodesic equation without any contribution from the Lorentz force. 
    
    To solve these equations, we first note that due to spherical symmetry, we can chose $\theta = \frac{\pi}{2}$ (equatorial plane) without the loss of generality. Then, the geodesic equations become,
    \begin{align}
        \ddot{t} + 2 \dot{r}\dot{t}\Phi^\prime - \e^{-2\Phi} \dot{r} e E(r) &= 0 \label{geoeq1}\\
        \ddot{r} + \dot{r}^2 \Psi^\prime + \e^{2\Phi - 2\Psi}\dot{t}^2 \Phi^\prime - \e^{-2\Psi}r\dot{\varphi}^2 - \e^{-2\Psi} \dot{t}e E(r) &= 0 \label{geoeq2}\\
        r\ddot{\varphi} + 2\dot{r}\dot{\varphi} &= 0 \label{geoeq3}
    \end{align}
    where, $E(r)$ is the static, radial electric field of the charged fluid. Further, due to metricity of the geometry, we also have,
    \begin{align}
        g_{\alpha\beta}\dot{x}^\alpha \dot{x}^\beta &= \mu \nonumber \\ \Rightarrow \ -\e^{2\Phi}\dot{t}^2 + \e^{2\Psi}\dot{r}^2 + r^2\dot{\varphi}^2 &= \mu \label{velnorm}
    \end{align}
    where, $\mu = 0\ {\rm and}\ -1$ correspond to null and timelike particles, respectively.

    Using equations \eqref{geoeq1} and \eqref{geoeq3}, one can find the conserved quantities for the motion of a test particle, given by,
    \begin{equation}\label{consquant}
        \e^{2\Phi}\dot{t} - e \mathcal{F} = \mathcal{E} \ ; \qquad r^2 \dot{\varphi} = L
    \end{equation}
    where, we have defined, $\mathcal{F}(r) = \int E(r) \df r$, $\mathcal{E}$ is the specific total energy and $L$ is the specific angular momentum of the particle. Using equation \eqref{consquant} in \eqref{velnorm}, we can write,
    \begin{equation}
        -\e^{-2\Phi}(\mathcal{E} + e\mathcal{F})^2 + \e^{2\Psi}\dot{r}^2 + \frac{L^2}{r^2} = \mu
    \end{equation}
	After rearranging the above equation, we get,
    \begin{equation}
        \e^{2\Psi}r^2 \dot{r}^2 = \mathcal{E}^2 \left[ \left(1 + \frac{e}{\mathcal{E}} \mathcal{F}  \right)^2 r^2 \e^{-2\Phi} + \frac{\mu}{\mathcal{E}^2}r^2 - \frac{L^2}{\mathcal{E}^2}\right]
    \end{equation}
    As is traditionally done, we define the impact parameter as, $b = \frac{L}{\mathcal{E}}$. Then, the motion of the test particles can be analysed easily by defining an effective potential given by,
    \begin{equation}\label{veff}
        V_{\rm eff} (r) = \left(1 + \frac{e}{\mathcal{E}} \mathcal{F}  \right)^2 r^2 \e^{-2\Phi} + \frac{\mu}{\mathcal{E}^2}r^2
    \end{equation}
    The effective potential follows the constraint, $V_{\rm eff} \ge b^2$. It is not surprising that in addition to the metric coefficients, the effective potential also depends on the electric field produced by the fluid sphere. This would mean that the motion of a (charged) test particle will be governed by both gravity and Lorentz force which is expected. For a neutral particle, the Lorentz force plays no part. However, the electric field (or charge) of the fluid does still affect the motion indirectly  through its contribution in the metric functions. Further, it is worth noting that the specific (not absolute) values of charge and energy of the test particles determine their trajectory. Only in the case of neutral null particles, the effective potential depends purely on the curvature and not at all on the properties of the particle.

    With the above expression for the effective potential, it is now straightforward to derive conditions for the trapping of geodesics. A trapped circular orbit exists if for some radius, $r_c < r_b$, we have,
    \begin{equation}
        \left.\frac{\df V_{\rm eff}}{\df r}\right|_{r_c} = 0
    \end{equation}
    using the expression in \eqref{veff}, this gives,
    \begin{equation}
        \left.\left(1 + \frac{e}{\mathcal{E}}\mathcal{F}\right)\ \e^{-2\Phi}\mathcal{E}^2 \left[ \left(1 + \frac{e}{\mathcal{E}}\mathcal{F}\right)\left( 1 - r \Phi^\prime \right) + \frac{e}{\mathcal{E}}\mathcal{F}^\prime r \right] \right|_{r_c}  = -\mu    
    \end{equation}
    The above condition reduces to different forms for different types of particles. Therefore, it is useful to distinguish these cases from each other. In table \ref{tab-trapcond}, we present each of these cases along with the corresponding condition for trapping.
    \begin{table}[H]
        \centering
        \begin{tabular}{l l l}
            \hline
            \hline
            \textbf{Particle} & & \textbf{Condition for Trapping} \\ [0.15ex]
            \hline
            & &\\ [-1.5ex]
             charged, massive ($e \ne 0,\ \mu = -1$) & & $\scalemath{0.8}{\left.\left(1 + \dfrac{e}{\mathcal{E}}\mathcal{F}\right)\ \e^{-2\Phi} \mathcal{E}^2\left[ \left(1 + \dfrac{e}{\mathcal{E}}\mathcal{F}\right)\left( 1 - r \Phi^\prime \right) + \dfrac{e}{\mathcal{E}}\mathcal{F}^\prime r \right]\right|_{r_c}   = 1}$\\ [1.5ex]
             charged, massless ($e \ne 0,\ \mu = 0$) & & $ \scalemath{0.8}{\left.\dfrac{e}{\mathcal{E}}\mathcal{F}\left( r \Phi^\prime - r\dfrac{\mathcal{F}^\prime}{\mathcal{F}} - 1 \right) + r\Phi^\prime\right|_{r_c} = 1}$ \\[1.5ex]
             neutral, massive ($e = 0,\ \mu = -1$) & & $\scalemath{0.8}{\left.\e^{-2\Phi} \mathcal{E}^2\left( 1 - r \Phi^\prime \right) \right|_{r_c} = 1}$\\ [1.5ex]
             neutral, massless ($e = 0,\ \mu = 0$) & & $\scalemath{0.8}{ \left.r \Phi^\prime\right|_{r_c} = 1}$\\ [1ex]
             \hline
             \hline
        \end{tabular}
        \caption{Conditions for the trapping of various types of geodesics in a static spherically symmetric charged space-time.}
        \label{tab-trapcond}
    \end{table}
    
    \subsection{Trapping Regions} \label{sec-paraexp}
\begin{figure}[H]
        \centering
        \includegraphics[width=0.92\linewidth]{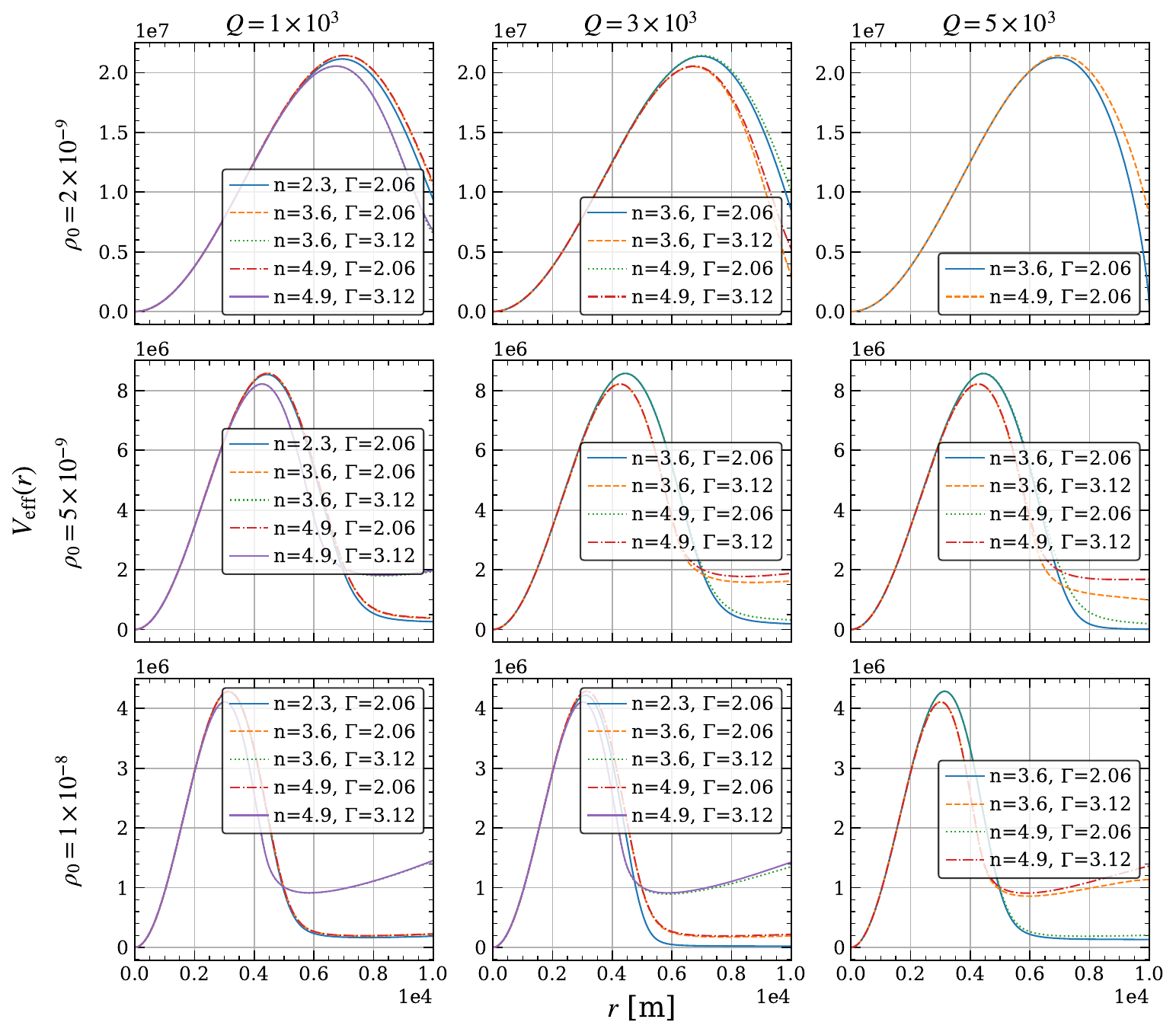}
        \caption{The effective potential, $V_{\rm eff}(r)$, for neutral, null particles as a function of the radial coordinate, $r$.}
        \label{fig-NMSprofile}
    \end{figure}

\begin{figure}[H]
        \centering
        \includegraphics[width=0.92\linewidth]{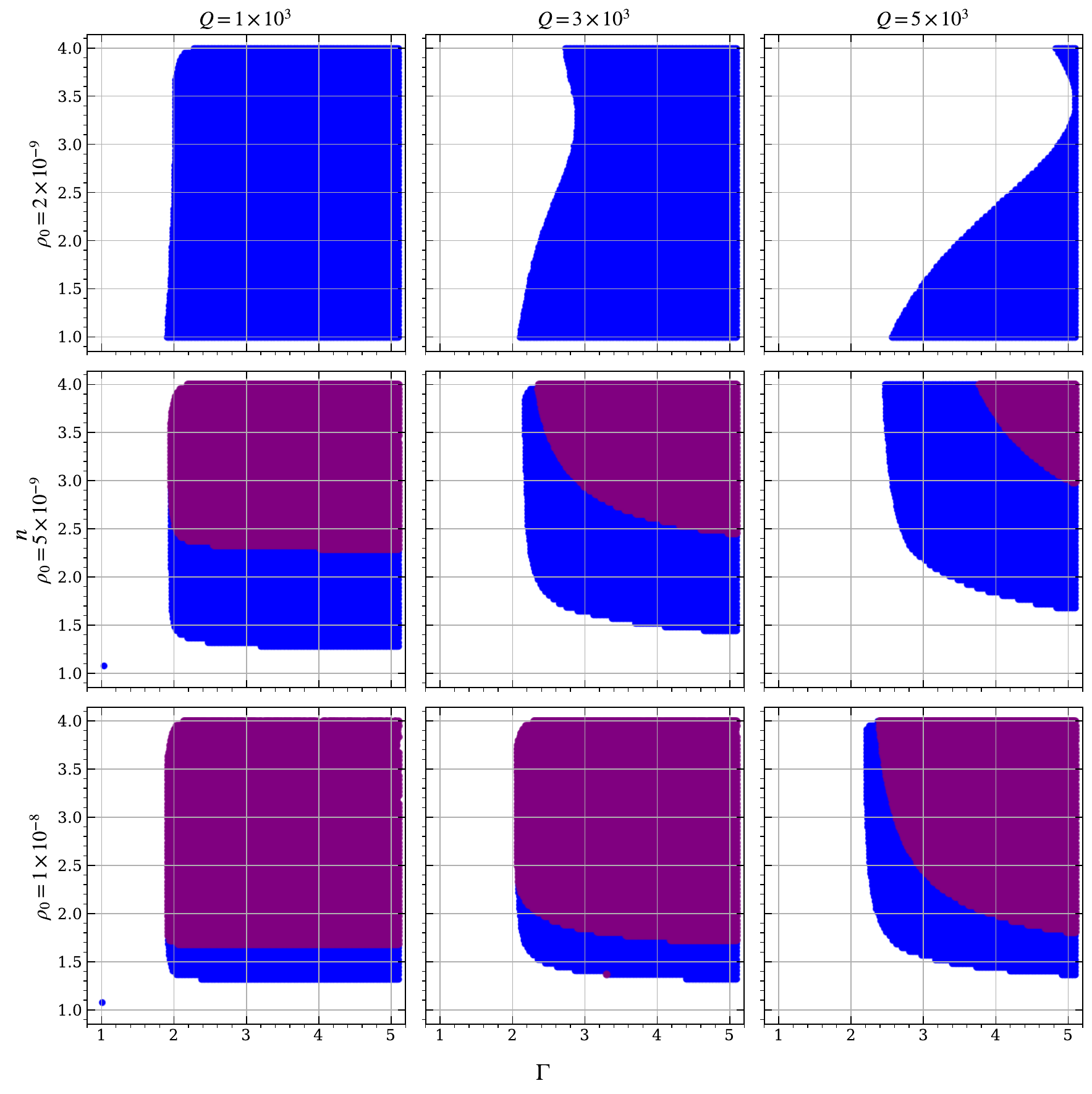}
        \caption{The trapping regions for neutral, null particles for various values of $\rho_0$ and $Q$. The region shaded \textit{blue} corresponds to existence of only stable orbits while \textit{purple} corresponds to both stable and unstable orbits.}
        \label{fig-NMSscatter}
    \end{figure}

\begin{figure}[H]
        \centering
        \includegraphics[width=0.92\linewidth]{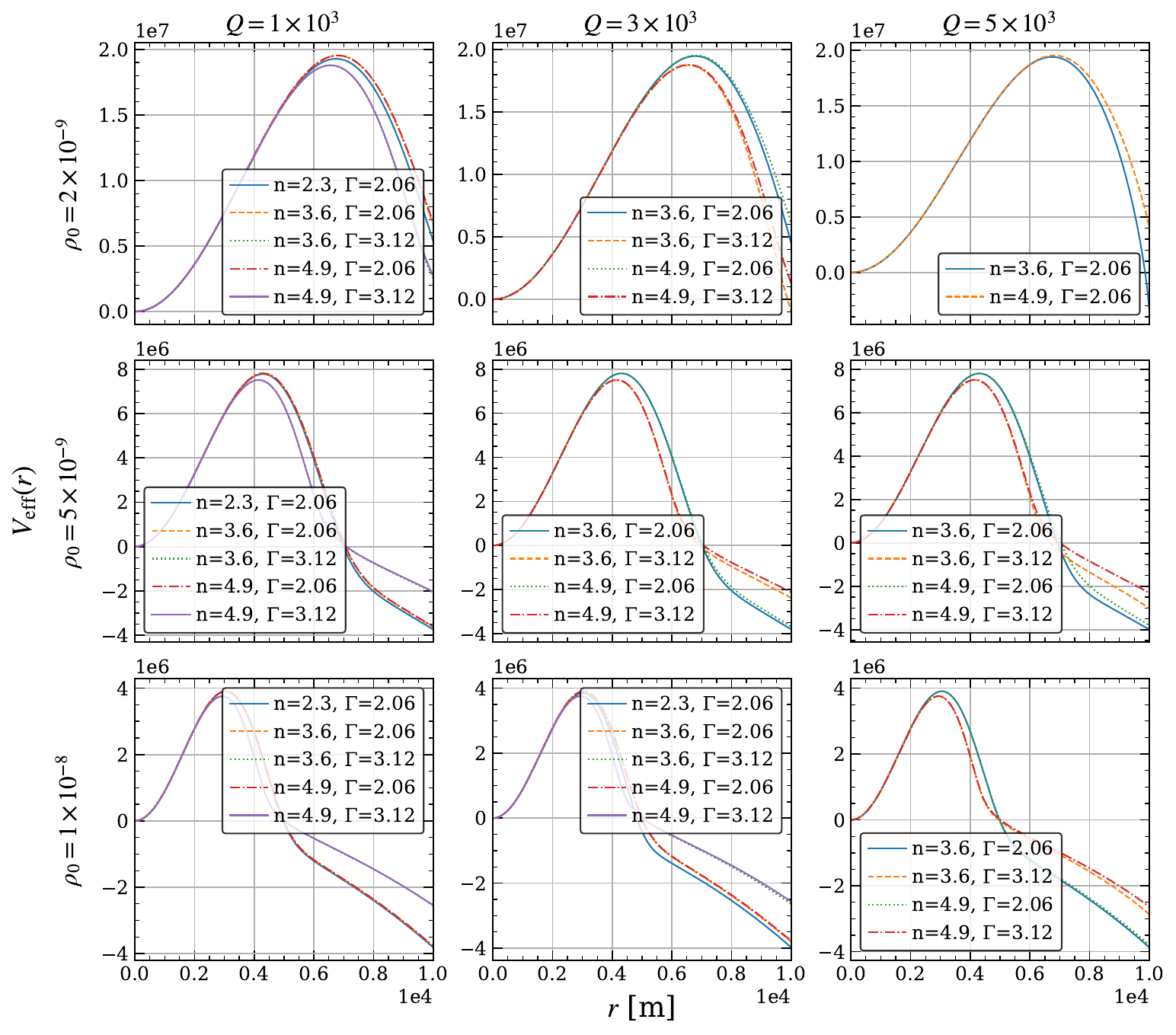}
        \caption{The effective potential, $V_{\rm eff}(r)$, for neutral, timelike particles as a function of the radial coordinate, $r$. We have used $\mathcal{E} = 5$.}
        \label{fig-NMSVprofile}
    \end{figure}

\begin{figure}[H]
        \centering
        \includegraphics[width=\linewidth]{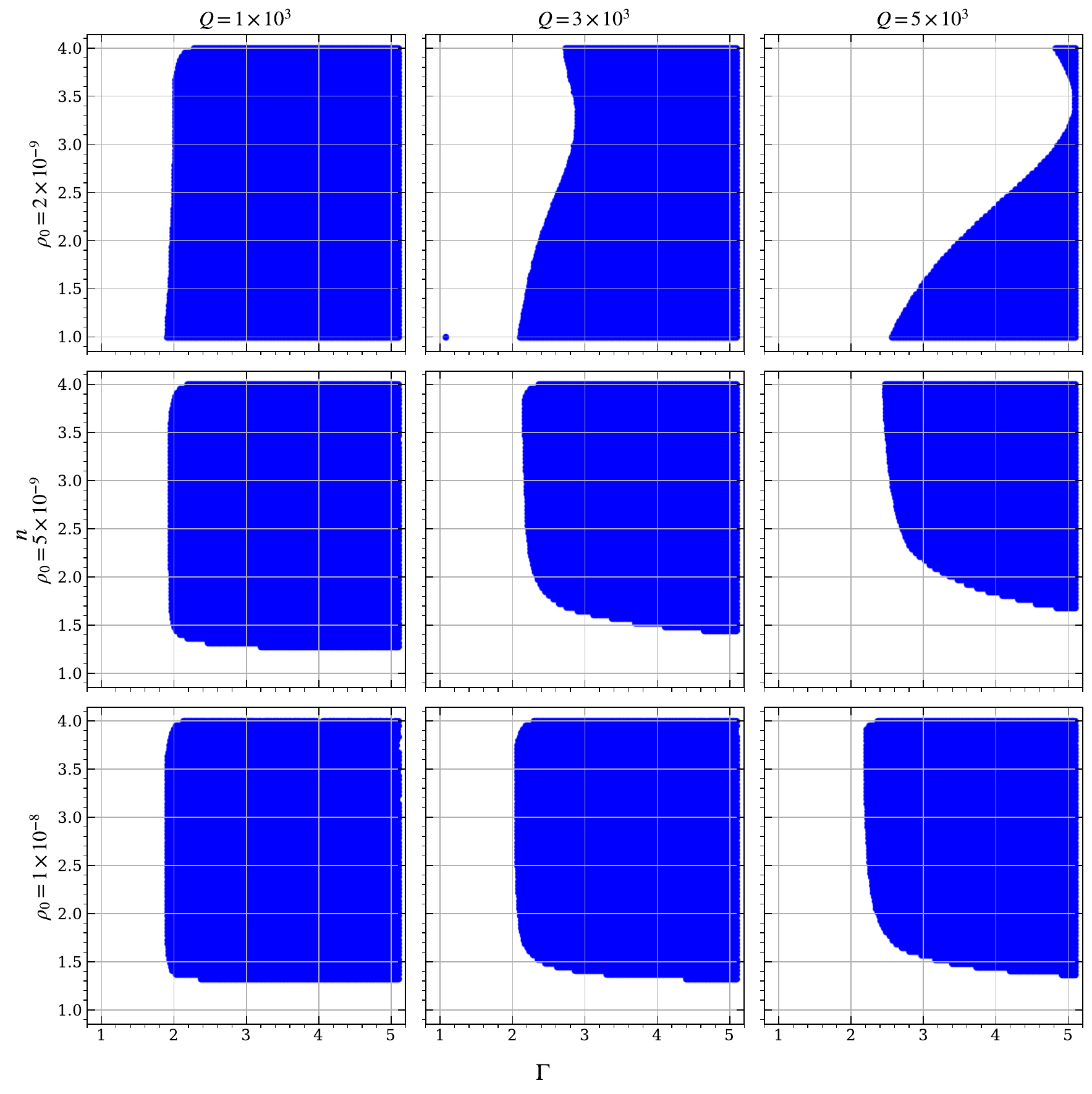}
        \caption{The trapping regions for neutral, timelike particles for various values of $\rho_0$ and $Q$. For the chosen range of $n,\ \Gamma,\ \rho_0,\ Q$, only stable orbits exist.}
        \label{fig-NMSVscatter}
    \end{figure}

\begin{figure}[H]
        \centering
        \includegraphics[width=\linewidth]{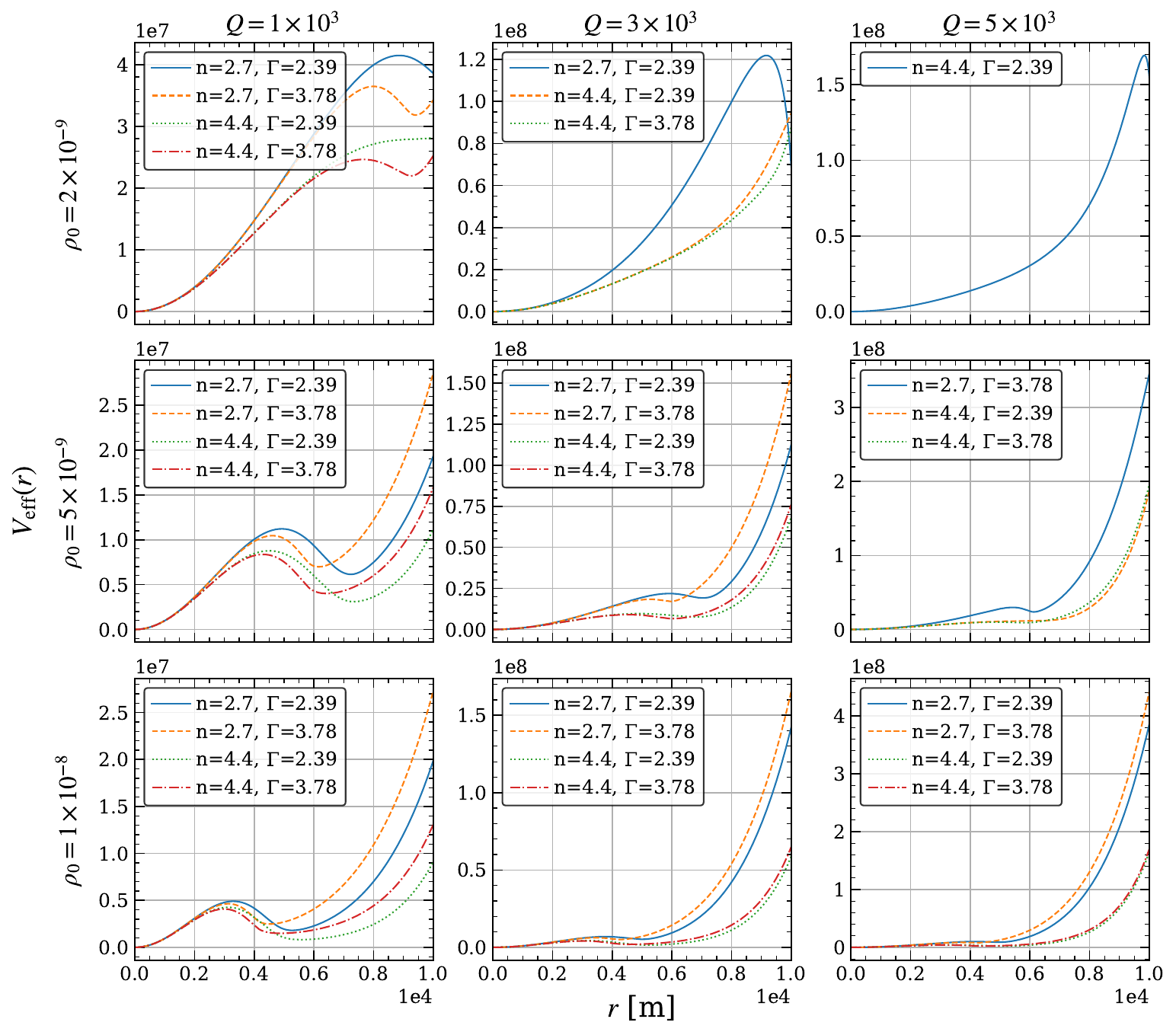}
        \caption{The effective potential, $V_{\rm eff}(r)$, for charged, null particles as a function of the radial coordinate, $r$. We have used, $\frac{e}{\mathcal{E}}=6$ as well as $\mathcal{F}(0) = 0$.}
        \label{fig-CMSprofile}
    \end{figure}

\begin{figure}[H]
        \centering
        \includegraphics[width=\linewidth]{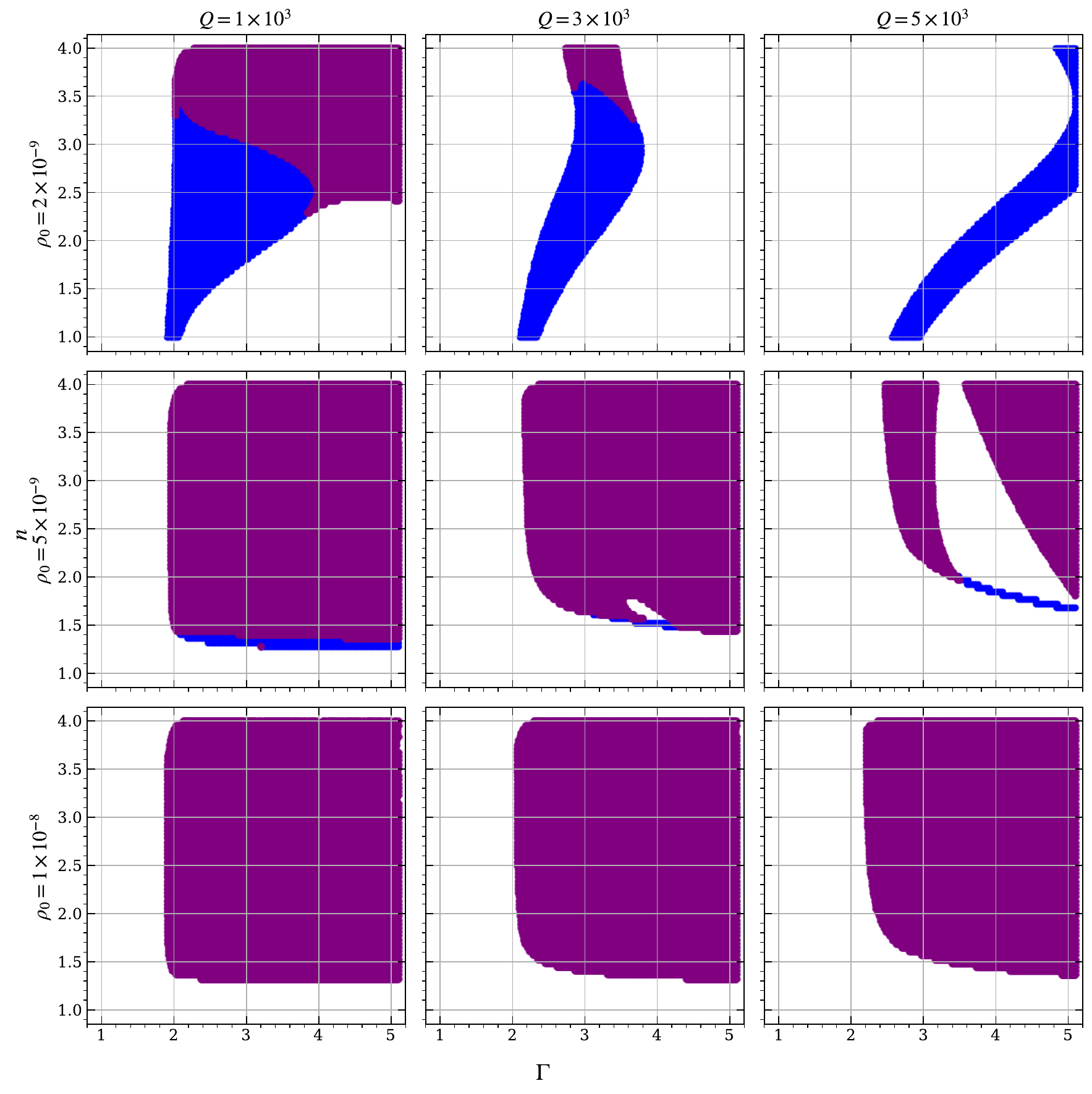}
        \caption{The trapping regions for charged, null particles for various values of $\rho_0$ and $Q$. We have used, $\frac{e}{\mathcal{E}}=6$. The region shaded \textit{blue} corresponds to existence of only stable orbits while \textit{purple} corresponds to both stable and unstable orbits.}
        \label{fig-CMSscatter}
    \end{figure}

\begin{figure}[H]
        \centering
        \includegraphics[width=\linewidth]{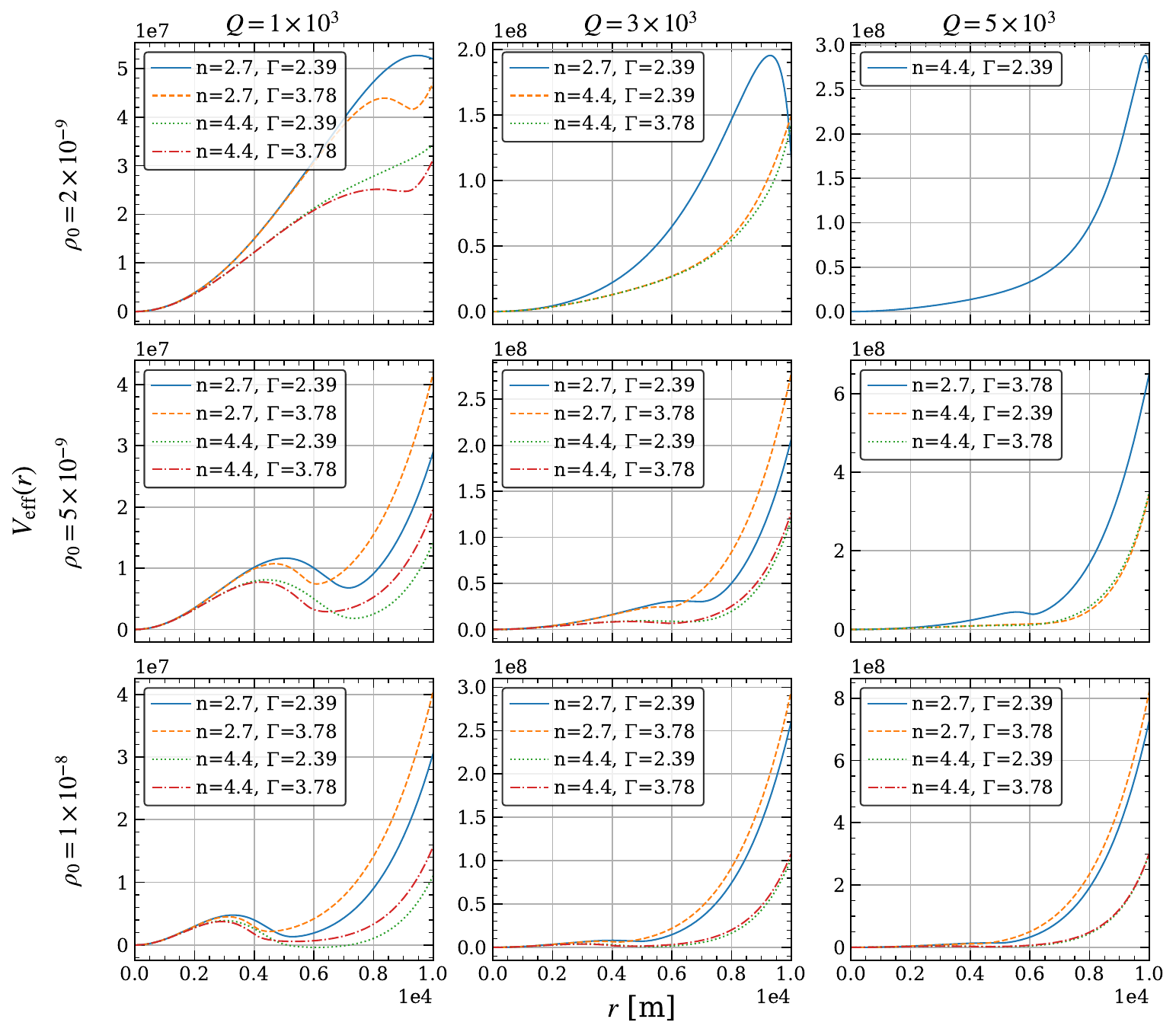}
        \caption{The effective potential, $V_{\rm eff}(r)$, for charged, timelike particles as a function of the radial coordinate, $r$. We have used, $\frac{1}{\mathcal{E}^2}=0.04$  and $\frac{e}{\mathcal{E}}=8.3$ as well as $\mathcal{F}(0) = 0$.}
        \label{fig-CMSVprofile}
    \end{figure}

\begin{figure}[H]
        \centering
        \includegraphics[width=0.92\linewidth]{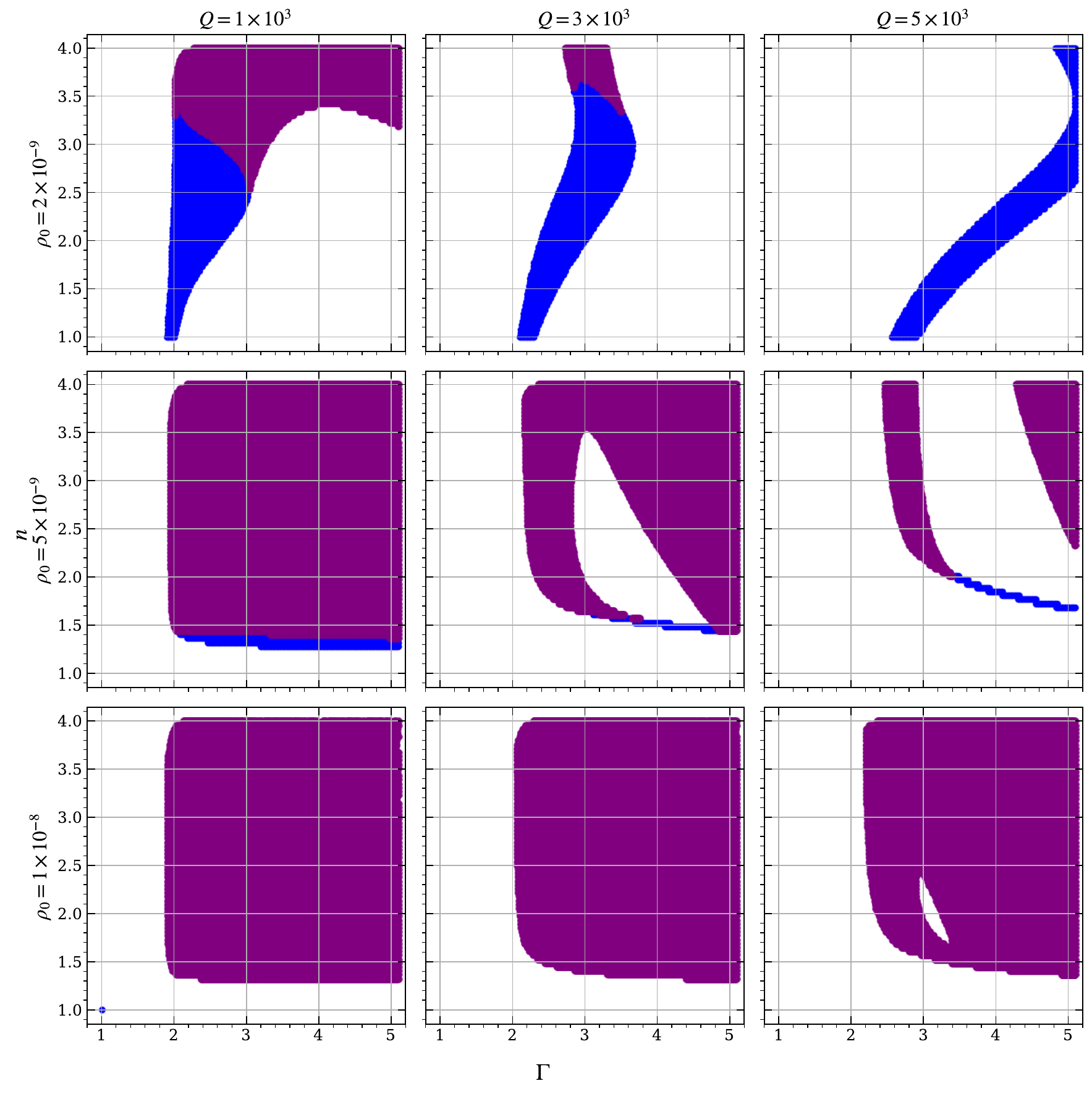}
        \caption{The trapping regions for charged, timelike particles for various values of $\rho_0$ and $Q$. We have used, $\frac{1}{\mathcal{E}^2}=0.04$  and $\frac{e}{\mathcal{E}}=8.3$. The region shaded \textit{blue} corresponds to existence of only stable orbits while \textit{purple} corresponds to both stable and unstable orbits.}
        \label{fig-CMSVscatter}
    \end{figure}

    With the expression for the effective potential and trapping conditions derived in the previous section, we now have everything necessary to check the existence of trapped orbits. We do this for all four types of particle listed in table \ref{tab-trapcond}. We  explore the $n$-$\Gamma$ parameter space and find regions that allow for trapped orbits, which we refer to as, `trapping regions'. Below, we present the plots for the effective potential (figures \ref{fig-NMSprofile}, \ref{fig-NMSVprofile}, \ref{fig-CMSprofile}, and \ref{fig-CMSVprofile}) and the trapping regions (figures \ref{fig-NMSscatter}, \ref{fig-NMSVscatter}, \ref{fig-CMSscatter}, and \ref{fig-CMSVscatter}) in all the four cases. All the plots are made with $\Lambda = 10^{-52}$ m$^{-2}$. To analyse the effect of changing $\Lambda,\ Q,\ r_b,\ \mathcal{E}$, and $\frac{e}{\mathcal{E}}$ on the trapping regions, we also generate animations sweeping through various values of these parameters\footnote{These animations could be found at this link: \url{https://github.com/astornelli/CPS}.}. The plots below can be considered as one of the frames in the corresponding animations corresponding to a fixed value of a given parameter that is being swept through. 
    
    We find that trapping regions exist for all types of particles for typical values of $\rho_0$, $\Lambda$, $Q$, and $r_b$. For neutral null particles, trapping regions become smaller as $Q$ or $\Lambda$ increase. For charged and/or massive particles, the energy per unit mass, $\mathcal{E}$ and the charge to energy ratio, $\frac{e}{\mathcal{E}}$ affects the trapping regions as well. We find that trapping regions do not exist for neutral timelike particles for $\mathcal{E}<1$ while increasing $\mathcal{E}$ beyond this does not have a significant effect. For charged timelike particles, increasing $\frac{e}{\mathcal{E}}$ makes the trapping regions smaller. For all four cases, the trapping regions become larger with increasing $r_b$.

    \subsubsection*{Neutral and Massless Particles}

    \subsubsection*{Neutral and Massive Particles}
    
    \subsubsection*{Charged and Massless Particles}
    
    \subsubsection*{Charged and Massive Particles}

    \section{Summary and Discussion} \label{sec-summary}
    In this work, we presented a preliminary analysis of static spherically symmetric charged polytropic configurations with a cosmological constant. We started with setting up the Einstein-Maxwell-$\Lambda$ field equations for a polytrope, $p \propto \rho^\Gamma$, and a power law radial charge distribution, $q \propto r^n$. Using this, we converted the generalised TOV equation into a `master equation' for the mass profile. All charged polytropic systems with a power law charge distribution can be analysed in an unified manner using this equation. We solved this equation numerically for typical values of the polytropic index, $\Gamma$, the exponent, $n$, central density, $\rho_0$ and total charge, $Q$. We found that the total mass, $m(r_b)$, obeys the generalised Buchdahl limit only for $Q \lesssim 10^4$. 

    Using these solutions, we analysed properties of the polytropic fluid and presented bounds on $n$ and $\Gamma$ that lead to physically acceptable configurations. We found that the models corresponding to $\rho_0 \sim 10^{-9}$ and $Q \sim 10^3$ fit the best to all the criteria for physicality (see figures \ref{fig-denprofile} - \ref{fig-sec}). Higher values of these parameters led to smaller regions of physical acceptability in the $n$-$\Gamma$ parameter space (figure \ref{fig-physaccept}). On the other hand, $\Lambda = 10^{-52}$ is very small to have any significant effect on the physical acceptability. However, values comparable to the central density would lead to violation of the strong energy condition. Then, we investigated the geometric properties of the space-time. We found that within the acceptable regions, both the metric functions were smooth and regular at the centre. We also showed that within this region, the spatial hypersurfaces exhibit a spherical geometry. We noted that varying the values of $\rho_0$ or $Q$ did not change the nature of geometry while large negative values of $\Lambda$ may. 
    
    Then, we turned our attention to the phenomenon of internal trapping of circular geodesics which has been previously studied mainly for neutral, massless particles in uncharged polytropic configurations \cite{stuchlik2017}. Here, we included the massive and charged cases as well as included a charge into the fluid. Using the effective potential for these cases, we showed that trapping is allowed for all types of particles for a broad range of $n$ and $\Gamma$. We found that higher values of $Q$ reduced the possibility of trapping while higher values of $r_b$ increased it. On the other hand, the cosmological constant enters the geodesic equations only via the metric function, $\e^{2\Phi}$. For the currently accepted astrophysical value of $10^{-52}$ m$^{-2}$, it has negligible effect on geodesics. As was the case with physical acceptability, for $\Lambda$ to become relevant for geodesic trapping, it must be comparable to $\rho_0$. In such cases, the cosmological constant leads to even smaller trapping regions. 

    We noted that, unlike in the case of neutral, massless particles, the effective potential for massive/charged particles also depends on properties such as specific energy, $\mathcal{E}$, and specific charge, $e$. For the case of neutral, massive particles, we found that trapped geodesics exist only for $\mathcal{E} > 1$. Increasing $\mathcal{E}$ beyond this does not have any qualitative effect on the trapping regions. This makes sense since from table \ref{tab-trapcond}, we can see that $\mathcal{E}$ must be greater than unity for the trapping condition to be satisfied since other terms are smaller than unity. For charged, massless particles, trapping takes place only for $\frac{e}{\mathcal{E}} \lesssim 10$ while decreasing the value of this ratio leads to larger trapping regions. For charged, massive particles, both $\mathcal{E}$ and $e$ play a role where their effect on the trapped regions remains qualitatively the same as in the other cases mentioned above.
    
    It is important to note here that in reality, trajectories of the test particles will also depend on additional physical effects such as radiation backreaction (for charged particles) and scattering. Therefore, our results here should be considered only a first step towards a more comprehensive analysis of trapped orbits in charged polytropic configurations. Moreover, charged, massless particles have not been observed in nature but do exist in various contexts in the literature \cite{kazinski2002,azzurli2014,lech1,lech2,lech3,fairoos2017,moitra2024}. Nevertheless, the possibility of trapped orbits means that this work might prove relevant for the study of neutrino trapping \cite{stuch2012,stuch2011,stuch2021} and black hole mimickers \cite{saida2015,johnson2018, stuchlik2017}.
    
    Finally, we remark on other possible constructions of static charged fluid spheres with $\Lambda$. One could construct models where, instead of a power law charge distribution, one assumes that the charge density is proportional to the matter density as done in \cite{ray2003,arbanil2013,arbanil2017}. This would fall into class Ic under the scheme described in section \ref{sec-efesol}. Other classes of solutions could also be derived. A generalised relativistic Lane-Emden type system with charge and $\Lambda$ can also be derived in a similar manner as \cite{picanco2004,stuch2016}. In addition to these, a direct follow up to the current work may include stability analysis and anisotropic extensions of our models \cite{bayin1982,herrera2013,sajadi2023}. We keep these issues, particularly the last two, for future work.

    \section*{Acknowledgements}
    We thank Shalini Ganguly and James Ripple for their help with an earlier, unpublished version of this work as well as with setting up the preliminary code for numerical integrations performed in the current version. The code and animations can be found at \url{https://github.com/astornelli/CPS}. We also appreciate inputs from one of the anonymous referees for the said earlier version. AA acknowledges partial support by an internal grant from St. Mary's College of Maryland.


\end{document}